\documentstyle[prb,aps,preprint,tighten,epsfig]{revtex}
\begin{document}
\title {Vortices in coupled planes with columnar disorder and
bosonic ladders}
\author{E. Orignac and T. Giamarchi}
\address{Laboratoire de Physique des Solides, Universit{\'e} Paris--Sud,
                  B{\^a}timent 510, 91405 Orsay, France \cite{junk} }
\maketitle
\begin{abstract}
The phase diagram of two  bosonic chains coupled by interchain hopping
and
interchain boson boson repulsion is investigated by means of
bosonization
and renormalization group techniques.
It comprises two types of charge
density waves, a conventional superfluid phase and a superfluid phase
made of condensed boson pairs. The effect of a random potential
on that system is
also derived . Compared with the one chain case, interchain hopping
strongly stabilizes the conventional superfluid phase with respect
to Anderson localization. By contrast, the  superfluid phase made of
condensed boson pairs is much
more unstable with
respect to Anderson localization than the superfluid phase of the one
chain
system.
 An exact  mapping of  the two strongly
correlated bosonic chains onto two coupled vortex planes
gives us the phase diagram of the latter system without any
further calculation. 
 The vortex
system exhibits two types of vortex lattice
phase(one with entanglement and another one without entanglement) ,
 a normal phase and a phase with bound pairs of vortices.
The results for Anderson localization of the bosons
  allow us to study vortex pinning by columnar defects or twin boundaries 
and obtain the correlation
length of the pinned vortex lattice.
We show that interplane vortex hopping strongly reduces vortex
pinning by columnar defects but not by twin boundaries.
 We obtain the melting temperature of the entangled solid.
 We also derive the critical current by modified Larkin-Ovchinnikov
arguments and find a very sharp decrease of critical currents when the
system moves into the entangled phase in the case of columnar pins.
For strong columnar pins, there is a transition into a pinned solid phase
with dislocations. We argue that two multicritical points may exist in
the phase diagram.
\end{abstract}
\section{Introduction}

Localization effects in quantum interacting systems remain, even now an
extremely challenging problem. For fermionic systems, although
interactions are necessary to describe a certain number of experimental
situations, the non-interacting case allows for a reasonable starting
point in most situations ,
allowing to deal with the complications due to
disorder\cite{lee_mit_long}.
For bosonic systems, the theoretical situation is much less favorable since
repulsive interactions have to be included from the very beginning.
Despite its theoretical intricacy, the problem of disordered bosons is
of importance for
a variety of experimental situations such as ${}^4$He in aerogels or
Vycor\cite{chan_vycor_localization}, granular superconductors
\cite{gerber_granular}, disordered superconducting
films\cite{lee_MoC_localization,hebard_InO_localization,paalanen_InO_localization}and
disordered Josephson
junctions\cite{vanoudenaarden_josephson_mott,vanoudenaarden_josephson_localization}.
More recently this problem has also
known a renewed interest, both theoretical and
experimental, due to the advent of High Temperature
superconductors\cite{blatter_vortex_review}. In these systems, thermal
fluctuations can be strong enough to unpin the Abrikosov Lattice from
impurities or even melt it
\cite{gammel_fusion_vortex,charalambous_melting_rc}
thus restoring  energy dissipation in the
system. Finding a way to pin the
vortex lattice more efficiently has therefore
become a problem of technological
importance\cite{ledoussal_nelson_splay} for applications of HTSC that
also arises important
questions in the basic physics
of vortex systems. Columnar defects parallel to the magnetic field are
expected to greatly enhance the pinning of vortices.
 From the theoretical point of view, it has been shown that the problem
of vortex pinning on columnar defects is equivalent to the problem of
Anderson
localization in a random potential of two dimensional bosons with
repulsive
interactions\cite{nelson_vortex_liquid,nelson_columnar_long}. In that mapping,
superfluidity of the bosonic system
corresponds to a vortex
liquid phase  and restoration of energy dissipation whereas
localization corresponds to pinning of
vortices and absence of energy dissipation.

The necessity of keeping the repulsive interactions in disordered
bosonic systems
is due to the fact in the
non-interacting limit, the ground state  corresponds to all bosons
condensed in the lowest
energy eigenstate resulting in an unphysical
 collapse of the system. As a result,
one has to start with a non-disordered boson system with
\emph{repulsive} interactions, and consider the effect of randomness.
In a two or three dimensional system, this is an extremely hard
problem. One dimensional systems allow for a
controlled treatment of the interactions therefore improving the
accuracy of the subsequent treatment of disorder.
It is thus not surprising that the first analytical solution of
this problem \cite{giamarchi_loc} was found for the case of
one-dimensional dirty bosons.
A transition between a superfluid phase and a localized phase upon
increasing the repulsion was found. The existence of such a transition
received numerical confirmation \cite{scalettar_bosons}, and was
extended to the more experimentally relevant two dimensional situation
by beautiful scaling arguments \cite{fisher_bosons_scaling}.
Since then localization of bosons in two dimension has been confirmed by
a variety of numerical techniques
\cite{krauth_bosons_disorder,sorensen_bosons_disorder,%
makivic_trivedi_bosons,wallin_girvin_bosons}.

However despite intensive efforts a controlled microscopic theory of such
a transition in more than one dimension is still lacking, and one has
to resort to various mean field type approximations, the validity
of which is difficult to control \cite{pazmandi_boson_mf}.
To try to interpolate between the one-dimensional solution and
a higher dimensional one,
we thus study in this paper the model of a two leg bosonic
ladder. Such a system has the advantage of containing interchain
tunneling, allowing to mimic higher dimensional systems,
while
retaining the controlled treatment of the interactions possible in the
one dimensional world. One thus expects
such a study to give some clue about higher dimensional systems. For
fermionic or spin system the study of pure\cite{balents_2ch,schulz_2chains,strong_spinchains}
or disordered\cite{orignac_2chain_short,orignac_2chain_long,orignac_2spinchains} ladders has
been extremely fruitful in that
respect. The microscopic results for the
ladder can also serve as a
test-bed for  scaling arguments \cite{fisher_bosons_scaling}
that are expected to be valid in all space dimension.  Indeed,
the scaling relations where found to be compatible with the exact
exponents \cite{giamarchi_loc} found in the single chain
superfluid-insulator transition.
In addition one expects our study to be directly relevant for
experimental systems such as Josephson junction networks that are
practical realization of one-dimensional
bosons\cite{vanoudenaarden_josephson_mott,vanoudenaarden_josephson_localization},
but often with
more than one channel involved in charge transport.

The ladder situation becomes also specially important when dealing with
vortex systems in Type II
superconductor \cite{blatter_vortex_review}.  The
superfluid to insulator
transition  that occurs in the dirty
bosons system when interparticle repulsion is increased
results  in the equivalent vortex system in a transition from an unpinned
vortex liquid to a pinned vortex solid
as temperature is lowered. This pinned phase is known as ``Bose
Glass'' phase and its physical properties have been analyzed in
Ref.~\onlinecite{nelson_columnar_long} and have been the object of
intense experimental studies
\cite{jiang_bose_glass_ybco,budhani_bose_glass_tlbacacuo,samoilov_bose_glass}.

Treating such system analytically proves to be difficult and possible
only in the elastic limit where not dislocations are allowed in the
lattice structure \cite{giamarchi_columnar_variat}. Although it is
indeed the case for weak
disorder at low temperature, as for point like
disorder\cite{giamarchi_vortex_short,giamarchi_vortex_long,giamarchi_diagphas_prb}
where the
existence of a dislocation free glassy phase was
proven both analytically
\cite{giamarchi_vortex_long,carpentier_bglass_layered,kierfeld_bglass_layered,%
fisher_bragg_proof} and
numerically \cite{gingras_dislocations_numerics}
such a
restriction cannot allow to approach the melting transition nor the case
of strong disorder where dislocations obviously play a role.
It is thus interesting to study special geometry, where dislocations
could be treated exactly. This is for example the case for
vortices confined into planes parallel to the magnetic
field\cite{golub_layers},
corresponding to coupled chains of one dimensional bosons.
This situation may be experimentally achieved by putting a magnetic
field parallel to the CuO planes in superconductors such as YBCO or
BiSSCO
\cite{jakob_multilayers,li_superconducting_multilayers,%
triscone_multilayers}.
In the case of point like disorder this geometry allowed to confirm the
existence of the Bragg glass phase\cite{carpentier_bglass_layered,kierfeld_bglass_layered},
and to study the melting and strong
disorder case in a controlled way. However in these approaches tunneling
of vortices between planes was not treated. When included \cite{balents_smectic}
it was argued
that the periodic potential
induced by the layers would still lead to a smectic order of the
vortices, with even the possibility of supersolid phases between
a crystal at low temperature and the liquid (or smectic) at high
temperatures.

The ladder system allows for a detailed microscopic solution of such
systems, taking both  hopping, interactions and disorder into
account. It is thus possible to describe more realistically the possible
phases of a two dimensional system, and in particular to take into
account such effects as the entanglement of the vortices. Indeed
in the vortex language, boson tunneling
translates as the ability for bosons to overcome the potential barrier
that maintains them in the planes by gaining energy from
 thermal fluctuations and form
``superkinks'' \cite{nelson_columnar_long}.
Also, the perturbative renormalization group arguments
leading to the supersolid  phase also applied to the two planes vortex
system. Therefore, the existence of such phase can be tested
in the two plane system going beyond perturbation theory.

The plan of the paper is as follows:
In  section \ref{reminder+puresystem}, we  recall the bosonization
procedure for bosonic systems. Then,
we derive the bosonized Hamiltonian of two coupled bosonic
chains. This allows us to obtain the phase diagram of the two chain
bosonic system using renormalization group arguments.
In section \ref{boson_localization} we consider the effect of random
potential scattering. We obtain the phase diagram in the presence of
disorder as well as the boson localization lengths and discuss the
physical consequences.
In section \ref{mapping-vortices-bosons}, we consider first the
mapping of interacting vortices
confined in planes onto interacting bosons. We give a physical
interpretation of the quantities defined in the boson language in
terms of vortices.
We discuss the phase diagram of the vortex
system in the absence of columnar disorder. Then, we discuss the effect
of columnar disorder and compare with the one chain system. Finally,
we give a calculation of the critical
current of the coupled planes vortex system.
Conclusion can be found in section~\ref{sec:conc}. Some technical
details have been put in the appendixes.

\section{Pure boson ladder: phase diagram} \label{reminder+puresystem}

\subsection{Hamiltonian}\label{boson_no_disorder}

\begin{figure}
 \centerline{\epsfig{file=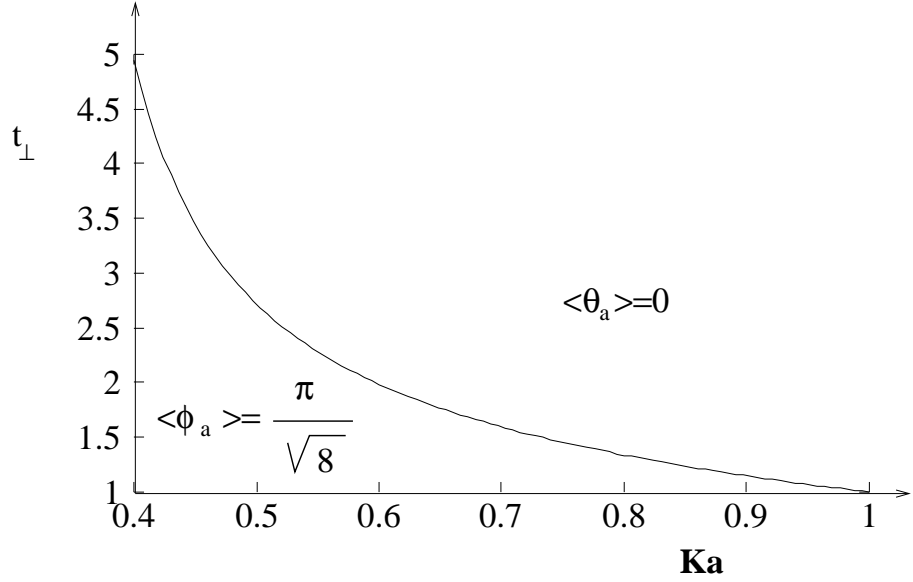,angle=-90,width=12cm}}
   \vspace{0.5cm}
\caption{The schematic phase diagram in the antisymmetric degrees of
freedom as a function of $K_a$ and $t_\perp$.
The line separates the phase with $\langle \phi_a^2 \rangle <\infty$
from the phase with $ \langle \theta_a^2 \rangle< \infty$. For $K_a>1$
the system cannot sustain a phase with $\langle \phi_a^2 \rangle
<\infty$. For $1/4 <K_a<1$ a nonzero $t_\perp$ is needed in order to
stabilize a phase with $\langle \theta_a^2 \rangle< \infty$. For
$K_a<1/4$, the only possible phase is the phase with $\langle \phi_a
\rangle=\frac{\pi}{\sqrt{8}}$}
\label{fig:phasediag}
\end{figure}

Let us consider the most general bosonic ladder with both interchain
hopping and interchain interactions. Its Hamiltonian is
\begin{eqnarray} \label{boson_2nd_quantized}
H & = & \int dx
\left[\frac{\nabla\psi^\dagger_{B1}.\nabla\psi_{B1}}{2m}
+\frac{\nabla\psi^\dagger_{B2}.\nabla\psi_{B2}}{2m}\right] +
\int dx dy U(x-y)\rho_{B1}(x)\rho_{B1}(y)\nonumber \\
& & +\int dx dy U(x-y)\rho_{B2}(x)\rho_{B2}(y)+
\int V\rho_{B1}(x)\rho_{B2}(x) dx \nonumber \\
& & +
\int t_\perp(
\psi^\dagger_{B1}(x)\psi_{B2}(x)+\psi^\dagger_{B2}(x)\psi_{B1}(x))dx
\end{eqnarray}
where $t_\perp$ is the interchain hopping, $U$ and $V$ are
respectively the intrachain and interchain interactions between the
bosons. Although we used for (\ref{boson_2nd_quantized}) a continuous
representation it of course also applies to discrete models (such as
e.g. the bosonic Hubbard model) provided the boson filling is
incommensurate with the lattice (for the case of commensurate filling
see Ref. \onlinecite{giamarchi_attract_1d} for a single chain or
Ref. \onlinecite{orignac_2spinchains} for a commensurate ladder).
Physically, one has $t_\perp<0$ and $V>0$.
To treat (\ref{boson_2nd_quantized}) is it convenient to use
an alternate boson representation of the original bosonic
degrees of freedom \cite{haldane_bosons}.
The salient points of this bosonization of bosons can be
found in Appendix~\ref{reminder} where the
single boson operator and the density operator are expressed as
(see Appendix~\ref{reminder})
\begin{eqnarray} \label{equ:basic}
\rho_{B}(x) & = & (\rho_0 +\frac{\partial_x
\phi}{\pi})\sum_{m=-\infty}^\infty e^{2\imath m(\phi(x)+\pi \rho_0 x)}
\nonumber \\
\psi_{B}(x) & \sim & e^{\imath \theta(x)} [\rho_B(x)]^{1/2} \sim
 e^{\imath \theta(x)}\sum_{m=-\infty}^\infty e^{2\imath
m(\phi(x)+\pi\rho_0 x)}
\end{eqnarray}
where $\rho_0$ is the average boson density,$\theta(x)=\pi
\int_{-\infty}^x dx' \Pi(x')$ and $\phi$ and $\Pi$
are conjugate operators and $a$ is a short distance cutoff.
The Hamiltonian (\ref{boson_2nd_quantized}) is bosonized by treating
separately the two chains. In (\ref{boson_2nd_quantized}), a continuum
Hamiltonian is used,
but the bosonization procedure would also work for a lattice
Hamiltonian. To do so, one introduces the two sets of
conjugated fields $\theta_1$, $\phi_1$ for chain 1 and
$\theta_2$, $\phi_2$
for chain two. In the absence of $t_\perp$ and $V$ the low energy
properties of each
chain are described by a Hamiltonian of the form (\ref{boson_formulas})
describing the sound waves which are the phonon modes
typical of a Bose superfluid.
All the interaction effects are hidden in the so called
Luttinger liquid parameters $u$ and $K$ (for each chain).
$u$ is the sound wave phase velocity whereas $K$ controls
the asymptotic decay of the correlation functions
(see appendix~\ref{reminder}). $u$ and $K$ can be extracted
directly out of physical quantities such as compressibility
and charge stiffness, as explained in Appendix~\ref{reminder},
regardless of the precise form of the interaction $U$.
For noninteracting bosons ($U\to 0$) $K\to \infty$, whereas
a local hard core repulsion would correspond to $K=1$. A longer range
interaction allows to reach all positive values of $K$.
Taking for simplicity identical chains one has $u_1=u_2=u$
and $K_1=K_2=K$.
The interchain boson-boson interactions $V$ and interchain hopping terms
$t_\perp$ are  expressed in terms of the fields
$\phi_1,\theta_1,\phi_2,\theta_2$ using (\ref{equ:basic}).
It is convenient  to use the symmetric and antisymmetric  linear
combinations:
\begin{eqnarray}\label{linear_combinations}
\theta_s=\frac{\theta_1+\theta_2}{\sqrt{2}},\theta_a=\frac{\theta_1-
\theta_2}{\sqrt{2}}\\
\phi_s=\frac{\phi_1+\phi_2}{\sqrt{2}},\phi_a=\frac{\phi_1-\phi_2}{\sqrt{
2}}
\end{eqnarray}
In this representation symmetric degrees of freedom
and antisymmetric ones decouple leading
to the following bosonized form for the Hamiltonian
(\ref{boson_2nd_quantized})
\begin{eqnarray} \label{bosonize_bosons}
H & = &H_{a}+H_{s}\nonumber\\
H_{s} & = & \int \frac{dx}{2\pi}\left[ u_{s} K_{s} (\pi \Pi_{s})^2 +
\frac{u_{s}}{K_{s}}(\partial_{x}\phi_{s})^2
\right] \nonumber\\
H_{a} & = & \int \frac{dx}{2\pi}\left[ u_{a} K_{a} (\pi \Pi_{a})^2
+\frac{u_{a}}{K_{a}}(\partial_{x}\phi_{a})^2
\right] +\frac{t_\perp}{\pi \alpha}\int dx \cos(\sqrt{2}\theta_{a})
+\frac{V\alpha}{(\pi\alpha)^2}\int dx \cos(\sqrt{8}\phi_{a})
\end{eqnarray}
The Luttinger liquid parameters for the symmetric and antisymmetric
modes are given by
\begin{eqnarray}  \label{parameters}
u_{s,a} &=& u\sqrt{1 \pm \frac{VK}{\pi u}} \\
K_{s,a} &=& \frac K {\sqrt{1 \pm \frac{VK}{\pi u}}}
\end{eqnarray}
where the upper sign is for the $s$ mode.
$H_s$ has the same form as (\ref{boson_formulas}) so that correlation
functions of the symmetric degrees of freedom are easily obtained
using (\ref{correlations}).
On the other hand, the antisymmetric degrees of freedom are more
difficult to treat. To do so one can exploit the analogy of $H_a$
with the Hamiltonian encountered for coupled spin chain systems
\cite{schulz_spins,strong_spinchains,orignac_2chain_long}. $H_a$ has two
different gapped phases. When $V$ is the most
relevant field, $\phi_a$ acquires a mean value that
minimizes the ground state energy (i.e. $\langle \phi_a \rangle=\frac
\pi 8$ for $V>0$),
whereas the correlation functions
of the exponentials of $\theta_a$ all decay exponentially fast.
When $t_\perp$ is the most relevant field
the situation is reversed: $\theta_a$ has a mean value that minimizes
the ground state energy (i. e. $\langle \theta_a \rangle=0$) and all
exponentials
of $\phi_a$ decay exponentially.
In the first case, the system is dominated by interchain repulsion
and the density fluctuations between the two chains are frozen. In the
second case interchain hopping dominates corresponding to the standard
Josephson coupling and assuring that the superconducting phase is
identical ($\langle \theta_a \rangle=0$) between the two chains.
The boundary $K_a^c$ between these two regimes depends on the values of
$V$ and $t_\perp$. A way to estimate the boundary is to consider
that the most relevant field is the one with the shortest correlation
length. Using the standard formula for the correlation length of
sine-Gordon time equations (see appendix \ref{reminder}) one obtains
for the values at the boundary (for small $V,t_\perp$)
\begin{equation}  \label{eq:boundary}
\frac{\log(V)}{\log(t_\perp)} = \frac{2-2K_a^c}{2-1/2K_a^c}
\end{equation}
For $K=1/2$ the transition occurs when $V= t_\perp$.
Let us not that since $\phi$ and $\theta$ are conjugate fields it is not
possible to get simultaneous order in \emph{both} fields even if at the
perturbative level both operators in the Hamiltonian can be relevant
simultaneously. The schematic phase diagram is shown on
Figure~\ref{fig:phasediag}.
\subsection{Correlation functions and phase diagram}
\begin{figure}

 \centerline{\epsfig{file=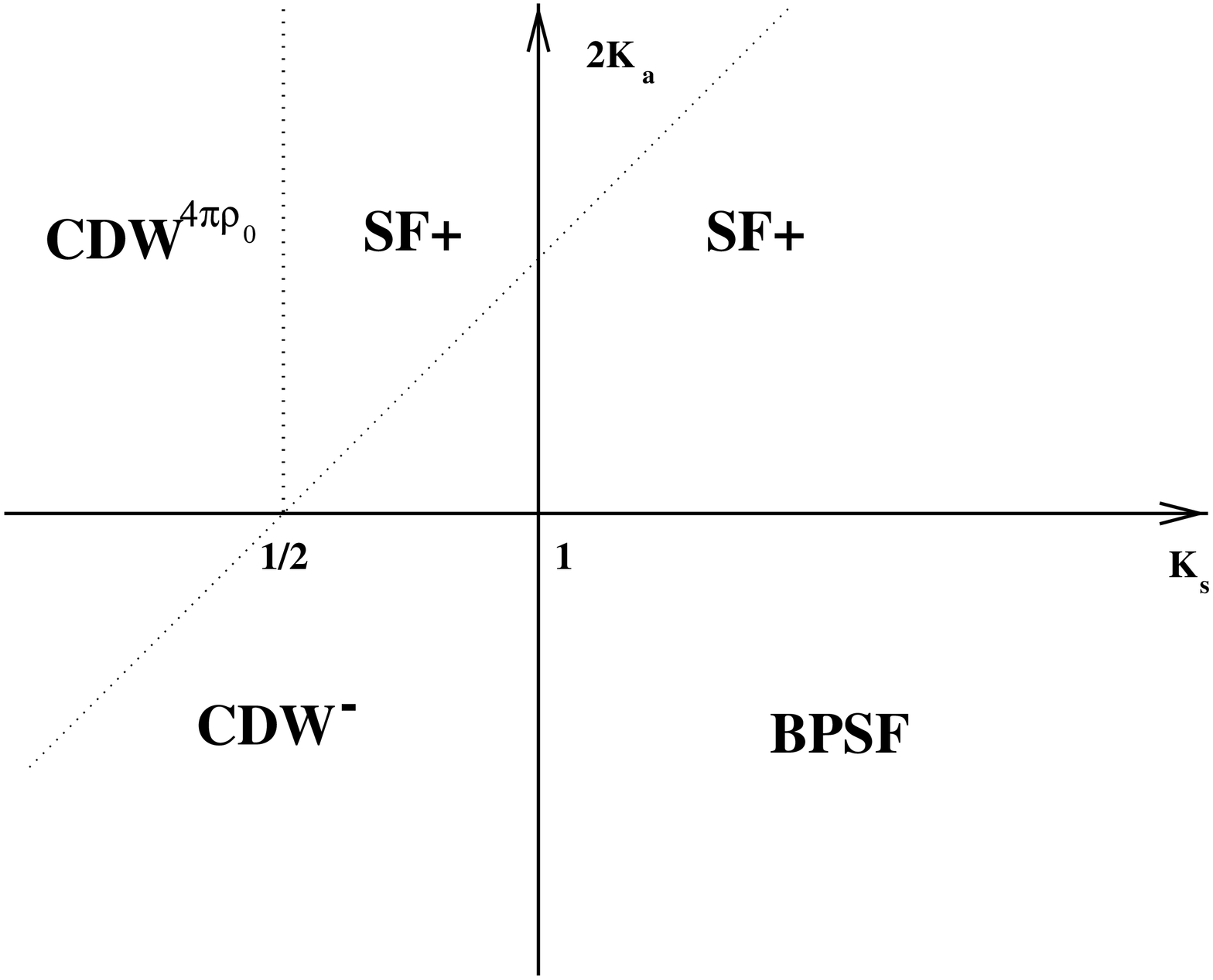,angle=-90,width=12cm}}
   \vspace{0.5cm}
\caption{The phase diagram of the  two chain bosonic system without
disorder.
The dashed line separates the domain of dominant CDW$^{4\pi\rho_0}$ from
the domain of dominant SF+ fluctuations. The separation between dominant
CDW- fluctuations and dominant BPSF fluctuations is at $K_s=1$. the
dotted line corresponds to weakly coupled bosonic chains.}
\label{fig:pure_bosonic}
\end{figure}

As a consequence of the existence of the gap in the antisymmetric
degrees
of freedom, most of the operators have exponentially decaying
fluctuations,and the associated susceptibilities remain finite.
Some operators however have algebraic correlations and diverging
associated susceptibilities. These are the following
\begin{eqnarray}
O_{SF+} & =
&\psi_{B1}+\psi_{B2}=e^{\imath\frac{\theta_s}{\sqrt{2}}}
\cos(\frac{\theta_a}{\sqrt{2}})
\label{superfluid}\\
O_{CDW^{4\pi\rho_0}}& = &\rho_{B1,4\pi\rho_0}+\rho_{B2,4\pi\rho_0}\sim
e^{\imath\sqrt{8}\phi_s} \label{ODC_4fermi} \\
O_{CDW-}& = &
\rho_{B1,2\pi\rho_0}-\rho_{B2,2\pi\rho_0}=\frac{e^{\imath\sqrt{2}\phi_s}
}{2\pi\alpha}
\sin(\sqrt{2}\phi_a)\label{ODC_moins} \\
O_{BPSF}& = & \psi_{B1}(x)\psi_{B2}(x) \sim e^{\imath
\sqrt{2}\theta_s} \label{boson_pairs}
\end{eqnarray}
The $CDW$ order parameters describe ``crystalline'' type phases, whereas
the $SF$ ones correspond to superfluid phases.
The $O_{CDW-}$ operator is the order parameter of an antisymmetric
charge density wave. In that phase, there is a charge density wave on
each charge of the ladder. These charge density waves have the same
$2\pi\rho_0$ wavevector and are dephased so that a minimum of the
density on
one leg of the ladder corresponds to a maximum of the density on the
other leg. Since we consider mainly interchain repulsion $V>0$ we
confine ourselves to this order parameter. For an attractive interchain
interaction one would have a corresponding $O_{CDW+}$ order parameter
(with $\cos(\sqrt{2}\phi_a)$ instead of $\sin(\sqrt{2}\phi_a)$)
corresponding to a phase for which the two $2\pi\rho_0$ CDW on each
chain would be in phase.
The $O_{CDW^{4\pi\rho_0}}$ operator is the order parameter of a
symmetric
charge density wave with $4\pi\rho_0$ wavevector. This
corresponds to bosons that form a charge density wave with a period half
the
one of the CDW-. The physical picture is that of bosons forming a CDW-
but hopping from one chain to the other thus minimizing the hopping
energy. This has the effect of halving the period and changing an
antisymmetric charge density wave into a symmetric one.
Strictly speaking, the order parameter for the $4\pi\rho_0$ charge
density wave is in fact:
\begin{equation}\label{ODC_4fermi_no_approximation}
O_{CDW^{4\pi\rho_0}}=e^{\imath \sqrt{8} \phi_s} \cos \sqrt{8} \phi_a
\end{equation}
and expression (\ref{ODC_4fermi}) results from the combination of the
term $V\cos \sqrt{8}\phi_a$ of
equation (\ref{bosonize_bosons})
with equation (\ref{ODC_4fermi_no_approximation}) in the perturbative
expansion of the correlation
functions.
The $O_{SF+}$ operator is the order parameter  of a superfluid phase in
which there is Bose condensation of the
bosons that occupy the bonding band. This is the normal superfluid
order parameter.
On the other hand
the $O_{BPSF}$ operator is the order parameter of a superfluid
phase in which pairs of
bosons that are formed along the rungs  present a Bose condensation.
In a lattice system, these pairs would be formed of a boson on $i$-th
site
of chain 1, and another boson on the $(i+1)$-th site of chain 2 for a
repulsive $V$ (on the same site for $V<0$). This implies
that the BPSF also contains
antisymmetric charge density wave (i. e. CDW-) fluctuations.
Which phase is realized depends on whether interchain hopping or
interchain interactions dominate.
The CDW$^{4\pi\rho_0}$ and SF+ phases are typical of a system dominated
by
interchain hopping whereas the CDW- and BPSF phase correspond
of a system dominated by interchain repulsion. Using
(\ref{bosonize_bosons})
one can obtain the generic phase diagram for the boson ladder.
The gapped fields, their mean value and the phase that obtains are
listed in table~\ref{table4}. The resulting phase diagram is shown
on Figure~\ref{fig:pure_bosonic}. For moderate intrachain repulsion
$K_s >1$, and only the superfluid phase is possible. If $V$ is small
with respect to $u/K$, $K_a=K+O(KV/u), K_s=K+O(KV/u)$ and one moves on a
special line of the phase diagram with $K_a\sim K_s$ (dotted line on
Figure~\ref{fig:pure_bosonic}). Note that, as can be seen from
(\ref{parameters}), increasing the small $q$ interchain repulsion
pushes paradoxically the system to a regime where the interchain
\emph{hopping} is more relevant and thus destabilizes the crystal.
This is due to the fact that now regions of high density
on one chain face regions of low density on the other making
it easier for the hopping to act.
Therefore, in a system of two coupled chains of
bosons with repulsive local interchain
interactions the BPSF does not exist (see Figure~\ref{fig:pure_bosonic}),
which is physically reasonable. Instead, one  has
a superfluid SF+ phase for $K>1/2$ and
a CDW-  phase for $K<1/2$ corresponding to a boson ``crystal''.
For attractive enough interchain interactions one could
expect to stabilize the BPSF phase.
Finally let us make the connection between the bosonic ladder and other
models. First using the standard equivalence between spins and hard core
bosons (see e.g. \cite{fisher_xxz}), the bosonic ladder can be mapped to
a spin ladder. In the absence of magnetic field for the spin system, one
would have a commensurate filling of one boson every two sites.
The massive phases (Haldane or singlet) for the spin system would thus
correspond to an insulating Mott phases for the bosonic ladder, where
bosons are frozen along the chains but able to hop between the two
chains. The Ising antiferromagnetic phase corresponds to the boson
``crystal''. The massless XY phases are the superconducting ones, the
BPSF phase being equivalent to the XY2 phase. Since such systems have
been studied in details both for the pure system and in the presence of
disorder we refer the reader to Ref. \onlinecite{orignac_2spinchains}
for a more detailed analysis of the physical properties.
Here we mainly focus on the incommensurate situation that can be mapped
to the spin ladder in the presence of a finite magnetic field, playing
the role for the bosonic ladder of the chemical potential.
Such bosonic ladders can also be mapped to the standard fermionic
ladder in the presence of an on-site attraction
\cite{orignac_2chain_long}.
In that case the electrons pair in bound states and single electron
hopping becomes irrelevant. The system becomes equivalent to pairs
of carriers hopping for site to site with an effective hopping of the
order of $t^2/|U|$ for a strong attraction $U$. The interchain
hopping can also be viewed in that case as the standard Josephson
coupling between two superconducting chains. More details can
be found in ref.~\onlinecite{orignac_2chain_long}.
\section{Effect of disorder}\label{boson_localization}
Let us now add disorder to the bosonic ladder.
In a continuum system, the only possible form of disorder is a random
chemical potential. In the case of a lattice system, another possible
form of disorder besides a random on site potential is random hopping
along the chains. Although these two types of disorder can lead in
principle to different physical phases when particle hole symmetry
exists (as is for example the case for spin chains
\cite{orignac_2spinchains,doty_xxz}), for an incommensurate density,
they lead to the same coupling in the continuum limit. Since we confine
our study of the bosonic ladder to such a case, we can restrict
ourselves to consider a random chemical potential.
The disorder is thus modeled as two uncorrelated random Gaussian
potentials in chains $1$ and $2$.
In the continuum limit, the coupling of the bosons with those random
potentials is:
\begin{equation}
\label{random_potential_for_bosons}
H_{\text{disorder}}=\int dx \left[V_1(x)\rho_{B1}(x) +
V_2(x)\rho_{B2}(x)\right]
\end{equation}
where $\overline{V_{p}(x)V_{p'}(x')}=D \delta_{p,p'}\delta(x-x')$.
 Upon bosonization the full Hamiltonian including the coupling with
disorder has the following  form
\begin{eqnarray}\label{boson_boson_random}
H & = & H_a + H_s + H_{\text{disorder}}\nonumber\\
H_{\text{disorder}} & = &\int dx
\sum_m \left[V^{2m\pi\rho_0}_{1}(x)e^{\imath m
(\sqrt{2}(\phi_{a}+\phi_{s})(x))}+V^{2m\pi\rho_0}_{2}(x)e^{\imath m
(\sqrt{2}(\phi_{a}-\phi_{s})(x))}+
\text{H. c.}\right]
\end{eqnarray}
where $H_a$,$H_s$ are given by  (\ref{bosonize_bosons}) and $\overline{
V_p^{2m\pi\rho_0}(x)V_{p'}^{2m\pi\rho_0}(x')}=D_m \delta(x-x')
\delta_{p,p'}$.
The
$q\sim 0$ part of the random potentials has been dropped since it
does not affects the gaps for small disorder and does not induce
localization
as discussed in Ref.~\onlinecite{orignac_2chain_long}. As usual the most
relevant coupling to disorder is given by the lowest harmonic
\cite{giamarchi_loc} so we retain for the moment only $V^{2\pi\rho_0}$
and $D_1 \sim D$. Higher harmonics can be taken into account in a
similar way using (\ref{equ:basic}).
The analysis of the phase diagram of (\ref{boson_boson_random}) proceeds
in a similar way than for spin ladders \cite{orignac_2spinchains},
with the important physical difference of a gapless symmetric sector
for the boson case. Disorder has different effects on
the phases
dominated by interchain hopping (i.e. with $\langle \theta_a \rangle=0
$)
corresponding to sectors I and II
of table \ref{table4} and the ones dominated by
 interchain repulsion (i.e. with $\langle \phi_a
\rangle=\frac{\pi}{\sqrt{8}} $) corresponding to sectors II and IV of
table
\ref{table4}.
In each case, two phases are in competition.
In the former case, there is a competition between a $4\pi\rho_0$ CDW
and a SF+ phase.
The $4\pi\rho_0$ CDW phase can gain energy by pinning on the
$4\pi\rho_0$
component of the random potential whereas in the SF+ phase, density
fluctuations are too small to permit gaining energy from the random
potential.
In the latter case, the competition is between the $2\pi\rho_0$ charge
density wave and the BPSF phase. The mechanism is similar, however
one expects that the stabilization of the CDW- is stronger than the
stabilization of the $4\pi\rho_0$ CDW.
Technically, the case of the competition of the CDW- and the BPSF
phases is simpler and is discussed first.
\subsection{System dominated by interchain repulsion}
When the system is dominated by interchain repulsion, the two phases in
competition
are the $CDW-$ and the BPSF phase.
The effective Hamiltonian describing the coupling with disorder is
obtained by replacing $\phi_a$ by its mean value in
(\ref{boson_boson_random}).
It has the form~:
\begin{equation}\label{simplified-bpsf}
H_{\text{effective}}=\int V_{\text{eff.}}(x) e^{\imath\sqrt{2}\phi_s}
\end{equation}
where $ V_{\text{eff.}}(x)=\mid \langle e^{\imath\sqrt{2}\phi_a}
\rangle \mid \imath (V_1(x)-V_2(x))$ and only the most relevant (in
the RG sense) terms
have been kept.
 From  (\ref{simplified-bpsf}) the renormalization group equation for
disorder  is
straightforwardly obtained in the form:
\begin{equation}\label{rg-for-bpsf+disorder}
\frac{dD}{dl}=(3-K_s)D
\end{equation}
The renormalization group equation for $K$ is~:
\begin{equation}\label{rg-for-K}
\frac{dK}{dl}=-C^2 D
\end{equation}
where $C$ is a non-universal prefactor.
In a similar way than for a single chain \cite{giamarchi_loc}
these RG equations have a fixed line on which disorder is irrelevant,
and a strong coupling disordered fixed point.
For small $D$, the renormalization of $K$ can be neglected and
(\ref{rg-for-bpsf+disorder}) is sufficient to determine whether
the system flows to the fixed line or to the strong coupling fixed
point.
According to (\ref{rg-for-bpsf+disorder}), the disorder is
relevant  for $K_{s}<3$ and irrelevant for $K_s>3$.
The strong coupling
fixed point is in this case an antisymmetric charge
density wave CDW- pinned on the $2\pi\rho_0$ component of
disorder\cite{giamarchi_loc,orignac_2chain_long}.
Therefore, the BPSF phase is
unstable with respect to
a very small amount of disorder unless $K_s>3$.
This  is due to the coupling of the random potential to the slowly
decaying CDW- fluctuations (see
equation  (\ref{ODC_moins})) that are subdominant in the BPSF phase .
When interchain density-density interactions dominate
the superfluid-insulator transition occurs thus for much less repulsive
interactions ($K=3$) than for the single chain case ($K=3/2$).
The situation is reminiscent of what happens in fermionic ladders
\cite{orignac_2chain_long}, and is due to the reduction of
quantum fluctuations by the freezing of the transverse degrees
of freedom. The localization length of the bosons in the pinned
phase is easily extracted from the RG equation as the
length for which the renormalized disorder is of order one.
It gives
\begin{equation}\label{pin-length-bos-int}
l_{\text{loc.,2 ch.}}\sim\left(\frac 1 D \right)^{\frac 1 {3-K_s}}
\end{equation}
and as can be expected is much shorter than the localization length
of a single chain
\begin{equation}\label{loc_length_1chain}
l_{\text{loc.,1ch.}}\sim\left(\frac 1 D\right)^{\frac 1 {3-2K_s}}
\end{equation}
Note that as for the fermion case \cite{orignac_2chain_long}
the change in localization length occurs through a change in exponent
and not as a simple multiplicative factor.
Since the Bose fluid that condenses is
made of bosonic pairs, to obtain the exponent of the correlation
function of the superfluid order parameter at the transition
one must consider
\begin{equation}
\langle T_\tau O_{BPSF}^\dagger(x,\tau)O_{BPSF}(0,0)\rangle
=\left(\frac 1 {x^2+(u\tau)^2}\right)^{\eta_c/2}
\end{equation}
as the superfluid order parameter correlation function.
That correlation function decays as $r^{-1/K_s}$ in the
superfluid phase and at the
transition point, $K_s=3$. Thus, at the transition point $\eta_c=1/3$.
It is interesting to remark that the same exponent at the transition
point obtains in the case of one bosonic chain \cite{giamarchi_loc}.
Using the RG one can also compute the conductivity in the case of a
charged
Bose fluid made for instance of Cooper pairs. This gives an indication
of the amount of dissipation
that is generated in the system upon introduction of disorder.
In a neutral boson system, conductivity measures the normal current that
would
be induced by a gradient in chemical potential.
The high frequency dependence of the a.c.  conductivity can be obtained
in the form :
\begin{equation}
\sigma_{\text{a. c.}}(\omega)=\omega^{K_s-4}
\end{equation}
using the methods exposed in Refs.
\onlinecite{giamarchi_loc,orignac_2chain_long}.
 From the a. c. conductivity, it is easy to deduce the temperature
dependence of
the d. c. conductivity. In the calculation of
Refs.~\onlinecite{giamarchi_loc,orignac_2chain_long}, frequency just
serves
in determining a cut-off length $\frac{u_s}{\omega}$. Similarly,
temperature provides a cutoff
length $\frac{\hbar u_s}{k_B T}$. This leads to the following expression
for the temperature dependence of the d. c. conductivity~:
\begin{equation}
\sigma_{\text{d.c.}}(T)=T^{2-K_s}
\end{equation}
For $T\ll T_{\text{pin.}}=\frac{u_s}{l_{\text{2ch.}}}$
or $\omega \ll T_{\text{pin.}}$, the conductivity goes to zero
exponentially fast.
\subsection{System dominated by interchain hopping}
\begin{figure}
 \centerline{\epsfig{file=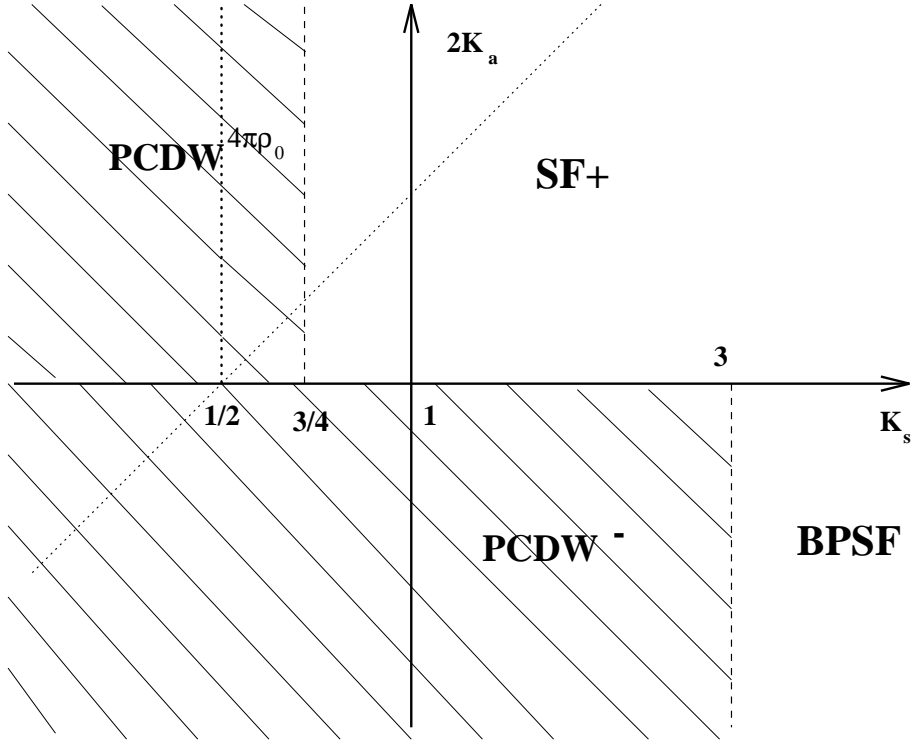,angle=-90,width=12cm}}
   \vspace{0.5cm}
\caption{The phase diagram of the  two chain bosonic system in the
presence of
a random potential. The dotted line corresponds to weakly coupled
bosonic chains. The hashed zones correspond to disordered phases in
which bosons are localized. These disordered phases are pinned charge
density waves. The stabilization of charge density waves by disorder
permits to have a disordered $4\pi\rho_0$ charge density wave for the
two weakly coupled chains.}
\label{fig:random_bosonic}
\end{figure}

In a system dominated by interchain hopping, the two phases in
competition
are the SF+ phase and the CDW$^{4\pi\rho_0}$.
Since in a system dominated by interchain hopping,
$\langle\theta_a\rangle=0$,  the massive
degrees of freedom in (\ref{boson_boson_random}) must be
integrated out in order to have a non trivial effective coupling to
disorder. Alternatively one can couple to a higher harmonic of the
random potential. Using the method of
Ref.~\onlinecite{orignac_2chain_long},the effective coupling
to disorder gives
\begin{equation}\label{disorder_in_SF+}
H_{\text{imp, eff.}}=\int dx
V_{\text{eff.}}(x)e^{\imath\sqrt{8}\phi_s} +\text{H. c.}
\end{equation}
where $\overline{V_{\text{eff.}}(x) V_{\text{eff.}}(x')}=C_1 D^2
\delta(x-x')$.
$C_1$ is a constant that comes from the integration of the gapped
modes\cite{orignac_2chain_long} .
Physically, this represents the coupling of the impurity potential
with the $4\pi\rho_0$ component of the density (see (\ref{ODC_4fermi})).
Note that the genuine $4\pi\rho_0$ component of the random potential is
coupled
to a $\cos(\sqrt{8} \phi_a)$, so that its contribution is a second
order term
$V^2D_2$. In this section we will thus use $\overline{D}$ for the
effective disorder.
The RG equation for disorder is obtained from
(\ref{disorder_in_SF+}) as
\begin{equation}\label{rg-for-disorder-in-sf+}
\frac{d\overline{D}}{dl}=(3-4K_s){\overline{D}}
\end{equation}
The impurity potential is relevant for $K_s< \frac 3 4$ and
irrelevant for $K_{s}>\frac{3}{4}$ .
Therefore, the SF+ phase
is stable with respect to Anderson localization for $K_s>\frac{3}{4}$,
corresponding to extremely repulsive interactions (this would need hard
core bosons plus a nearest neighbor repulsion).
For $K_s<\frac{3}{4}$, the randomness is relevant and the SF+ phase is
replaced by a $4\pi\rho_0$ CDW
pinned on the disorder.
The pinning length of the CDW, which is also the localization length
of the bosons, is given by~:
\begin{equation}\label{pin_length_bos-hop}
l_{\text{loc.,2 ch.}}\sim\left(\frac 1 {\overline{D}}
\right)^{\frac{1}{3-4K_s}}
\end{equation}
This time, the localization length is much \emph{larger} than the
localization length of the one chain system  (\ref{loc_length_1chain}),
indicating
that the two chain system is less localized than its one chain
counterpart.
In particular, a system of two
hard-core bosons chains (which has $K=1$) with interchain hopping
would be superfluid even in
the presence of a weak random potential, whereas its one chain
counterpart would have been already localized ($K_c = 3/2$).
The critical exponent of the superfluid order
parameter correlation function is obtained by the method of the
preceding
section. In the superfluid phase, the 2 point
correlation function of $SF+$ decay as $r^{-1/4K_{s}}$. Since the phase
boundary is at  $K_{s}=3/4$, the exponent of the superfluid order
parameter at the transition point is  $1/3$ as in the preceding section.
The high frequency dependence of the conductivity is computed as:
\begin{equation}
\sigma_{\text{a.c.}}(\omega)=\omega^{4K_s-4}
\end{equation}
using the methods of Refs.
\onlinecite{giamarchi_loc,orignac_2chain_long}.
Due to the equivalence of frequency and temperature one also has:
\begin{equation}
\sigma_{\text{d.c.}}(T)=T^{2-4K_s}
\end{equation}
As in the preceding section, high frequency conductivity refers to a.c.
conductivity at frequency  above the characteristic pinning frequency
$u_s/l_{\text{2ch.}}$. The full phase diagram in the presence of
disorder is represented on Fig.~\ref{fig:random_bosonic}
\subsection{Physical discussion}
Although most of the physical realization of bosonic ladders are likely
to come from classical systems (see next section), its study also
present direct interest for quantum systems.
Indeed recently practical realization of one dimensional bosonic systems
have been obtained, mostly based on Josephson junction
arrays\cite{vanoudenaarden_josephson_localization}.
It would be interesting to check in these systems various of the
predictions made here. The easiest to determine is certainly the
strong stabilization of the superfluid phase for the ladder in
comparison with the single chain case. For the pure system, for a
commensurability of one boson per site the Mott transition occurs
at $K=1$ (the hard core limit) for the ladder against $K=2$
\cite{haldane_bosons,giamarchi_umklapp_1d} for a single chain, i.e. much
less
repulsive interactions. In the presence of disorder a similar
stabilization occurs since the transition goes from $K=3/2$ for a single
chain to $K=3/4$ for the ladder. Maybe even easier to investigate
would be the bosonic ladder in the presence of a single barrier.
The properties of uniform disorder and a single impurity
\cite{kane_qwires_tunnel_lettre,kane_qwires_tunnel} can
be easily connected \cite{giamarchi_moriond} which allows from the
present
study to get the parameters for which the barrier would remain
transparent. For a single bosonic chain the barrier is not
effective for $K > 1$ whereas for the ladder the limit would be $K >
1/2$.
Another, albeit more theoretical interest of the bosonic ladder is
to provides another analytic system on which the
general scaling arguments \cite{fisher_bosons_scaling}
proposed for the superfluid-insulator transition can
be checked. Indeed at that time these arguments could
only be benchmarked on the only existing analytical
solution for a superfluid-disordered transition, namely the
$d=1$ single chain case \cite{giamarchi_loc}. Although since then
these scaling predictions have been found in agreement with various
numerical simulations
\cite{sorensen_bosons_disorder,krauth_bosons_disorder,%
makivic_trivedi_bosons}, it is still interesting to
verify them on models for which the exponents can be computed exactly.
The equality $z=d$ for the dynamical exponent clearly still holds
for the ladder since the system is still described by a Luttinger
Liquid. Similarly it is easy to check from the
RG equations that the compressibility
still remains finite at the transition, again in agreement with
the scaling prediction.
Finally,in Ref.~\onlinecite{fisher_bosons_scaling}, a bound
for the exponent of the superfluid order parameter at the
transition point has been derived by combining an exact inequality with
scaling.
When specifying to $z=d=1$,
that bound is  $\eta_c \le 1$. Since $\eta_c=1/3$ at the
transition, both in the case of the BPSF to insulator transition
and SF+ to insulator transition, the bound is indeed verified
for the ladder.
It would clearly be interesting to extend such an analysis to
ladders with more than two legs to complete the check.
Finally the two leg ladder prompts for the interesting question
of the existence of a \emph{universal} exponent for the superfluid
transition in one dimension $\eta = 1/3$, regardless of the number of
legs. Although a detailed check of this statement would need a complete
solution of the models with more than two legs, simple arguments given
in Appendix~\ref{ap:simple} suggest that this is indeed the case at
least in systems in which all modes but the symmetric one are
gapped. This is the case of all systems with an even number of legs.
However, this lets the possibility that the exponent could be different
from $1/3$ in systems with an odd number of chains.
An interesting question is whether the exponent is also $1/3$ for the
transition from the pinned charge density wave that occurs along the
line $3/4<K_s<3$, $K_a=1/2$ or if it depends of $K_s$.
It is difficult to answer this question in the framework of the
approximation of a disorder much smaller than the gaps, since such
approximation breaks down precisely in the vicinity of that line.

Let us note that of course the universality of the exponent holds here
for the case where the disorder is much weaker than all the gaps in the
system. The argument cannot be applied directly for a two dimensional
system.

\section{Coupled vortex planes problem} \label{mapping-vortices-bosons}
Although specially engineered quantum systems are necessary to have
coupled bosonic chains, such systems have a natural realization in the
classical world of vortices. Indeed, using a standard
mapping\cite{nelson_vortex_liquid,nelson_columnar_long} space time
trajectories of bosonic particles can be mapped onto a problem of
interacting flux lines. The
 disorder in the quantum problem is  time (i.e. $z$)
independent , corresponding in the vortex problem to columnar
defects.  Defects of that type can be artificially produced by
irradiation.
In the presence of such defects the vortex system is known to give rise
to a ``Bose glass'' phase corresponding to the localized phase of two
dimensional bosons.
On the other hand a very interesting situation occurs in a layered
system if the magnetic
field is aligned along the planes direction. In that case it has been
proposed\cite{balents_smectic} that the additional periodic potential
due to the planes
can confine the vortices and lead to a smectic
like state. If the confinement is strong enough one can view the
vortices as existing only in planes with repulsive interactions
between adjacent planes. Such coupled planes models are especially
interesting to study in their own since they allow a detailed
description of the melting transition for the pure system
as well as a study of the relevance of dislocations when disorder
is present\cite{carpentier_bglass_layered,kierfeld_bglass_layered,golub_layers}. However, in most of the studies of such systems
hopping of the vortices between planes was neglected.
When hopping is included, it was proposed\cite{balents_smectic} for the
pure system
that three phases would exist: a solid phase where no hopping is
present and the system crystallize due to interplane repulsion,
an intermediate supersolid phase where both the interchain repulsion
and the hopping are relevant, and a high temperature liquid phase.
Since the model does not allow for vortices to exist between the
planes the liquid phase also possesses a smectic order along the $z$
direction.
For an infinite number of planes such phases can only be determined
by looking at the relevance of the various operators (interchain
repulsion and interchain hopping).
For two coupled vortex planes we can use the results of the previous
sections to give precise answers to the nature of the
various phases. Since the perturbative study of two planes and
an infinite number of planes are identical we expect the results to be
also of relevance for the more general case. In addition one can study
the effect of columnar defects of twin boundary disorder
on the phases of the pure system. The more delicate case of
point like disorder will be considered elsewhere.
The particular case of two coupled planes of vortices could also be of
relevance to artificial layered structures where superconducting
structures can be separated by arbitrary heights of insulating
material\cite{jakob_multilayers,li_superconducting_multilayers,%
triscone_multilayers}.

\subsection{Mapping of the interacting vortices problem onto a bosonic
problem}
\subsubsection{Path integral reformulation of the interacting vortices
problem in terms of interacting bosons}
\begin{figure}
 \centerline{\epsfig{file=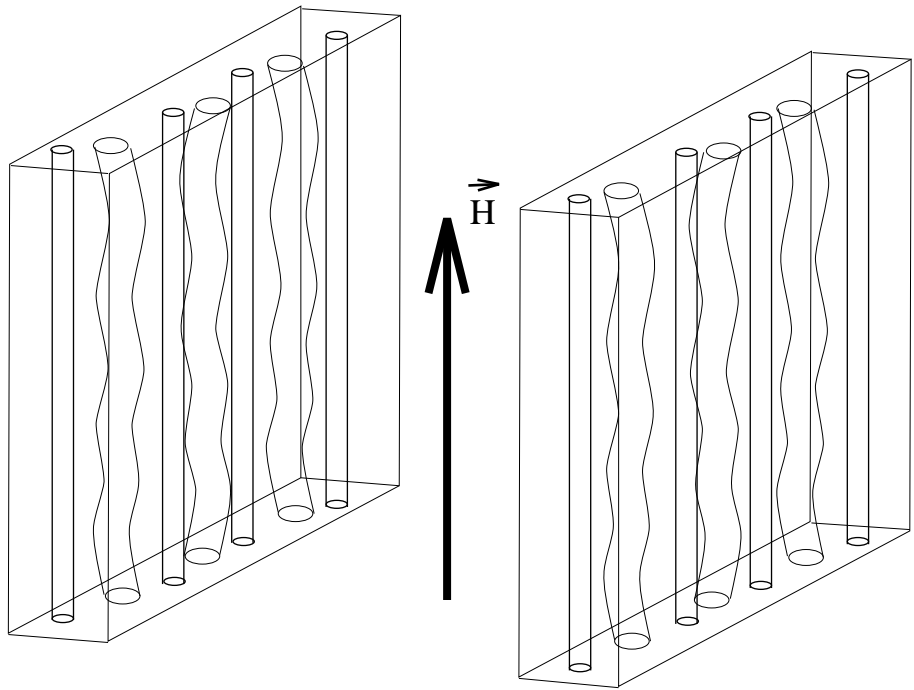,angle=-90,width=12cm}}
   \vspace{0.5cm}
\caption{The two planes vortex system. The vortices are confined in
the planes and can pin on columnar defects parallel to the magnetic
field.
In the situation represented on the figure, the temperature is low and
the vortices are confined to the planes}
\label{fig:boson-vortex}
\end{figure}

In this section, we consider a bilayered
superconductor in a magnetic field $H>H_{c_1}$ parallel to the bilayers.
We make the approximation that the bilayers are sufficiently far apart
so that the Josephson coupling of the bilayers can be neglected.
We suppose that the bilayers exert a strong enough attraction on the
vortices so that they remain confined in the planes. We also suppose
that there are columnar defects parallel to the planes and parallel to
the magnetic field as shown on Fig. \ref{fig:boson-vortex}.
Such system may be realized with copper oxide
superconductors with coupled planes (such as BiSSCO) or by epitaxial
growth of artificial structures
\cite{jakob_multilayers,li_superconducting_multilayers,triscone_multilayers}
with layers of two superconducting
materials, one of them being more efficient in confining vortices.

In the following, we design by $Oz$  the direction of the columns and
of the magnetic field, by  $Oy$
the direction perpendicular to the plane and $Ox$ the direction such
that
$(Ox,Oy,Oz)$ forms an orthonormal basis.
The position of a point on a flux line $i$ of coordinate $z$ along the
$Oz$ axis is given by ${\bf r}_i (z)=({r_i}_x(z),{r_i}_y(z))$.
The Hamiltonian for that system is then:
\begin{equation}
H=\int_0^\beta dz \left[ \sum_i  \left( \frac {\partial {{\bf
r}_i}}{\partial
z}\right)^2(z) + \frac 1 2 \sum_{i \neq j} U({\bf r}_i(z)-{\bf r}_j(z))
+ \sum_i (V({r_i}_y)+V({r_i}_y-2a)) +\sum_i v({\bf r}_i(z)) \right]
\end{equation}
$\beta$ being the height of the planes, $2a$ their distance, $U$ being
the vortex-vortex repulsive potential, $V$ the vortex plane
interaction and $v$ being the random potential induced by the columnar
defects.  We consider $v$ to be small, corresponding
to weak columnar disorder parallel to
the applied magnetic field. Although experimentally such a situation is
more difficult to achieve than strong pinning, it allows for an analytic
treatment of the problem and at least give
a qualitative understanding of collective pinning in such
systems. This collective pinning may be probed in high field
experiments.

The partition function of that system
\begin{equation}
Z=\int {\mathcal D}{\mathbf r}_i(z) \exp \left( -\frac H T \right)
\end{equation}
is identical\cite{nelson_vortex_liquid,nelson_columnar_long} to the one of a system of
interacting bosons at
temperature $\beta$ with a Planck's constant $T$.
Assuming that the attractive potential imposed on the vortices
by the planes is strong enough, the tight
binding approximation can be used to simplify the boson Hamiltonian
\cite{nelson_columnar_long}.
The calculations can be found in appendix \ref{tight-binding}.
The resulting Hamiltonian
is identical to (\ref{boson_2nd_quantized}), i.e.
two bosonic \emph{chains}, with interchain
and intrachain boson-boson interactions, interchain hopping and a
random intrachain potential.
Therefore,  we only have to translate the results already derived  for
the bosonic system in section \ref{boson_no_disorder} in the vortex
language.

\subsubsection{Interpretation of the bosonized vortex Hamiltonian in
terms of vortices}
In the case of a single chain, the field $\phi(x)$ has a simple
interpretation in terms of flux lines.
If
$u(x)$ represent the displacements of the vortex lattice in the
elastic theory , one has:
$\phi(x)=-\pi \rho_0 u(x)$ by identifying the expressions for the
$q\sim 0$ component of the density.
As a consequence, the action of the bosonized theory identifies with
the elastic Hamiltonian of the vortex lattice.
 The compressibility $\chi$  and  the charge stiffness ${\mathcal D}$
of the bosonic system defined respectively in equations
Eqs. (\ref{compressibility-def}) and (\ref{stiffness-def})  have also a
simple physical interpretation in terms of vortices.
$\chi^{-1}$ is proportional to the bulk modulus of the
vortex system.
$\chi^{-1}$ is also proportional to the magnetic susceptibility of the
system since the
vortex number is proportional to the
magnetic field.
$1/{\mathcal D}$ is proportional the shear modulus of the vortex
lattice.  This
results from writing the action of the bosonized theory (which is also
the Hamiltonian for the vortex lattice elastic theory) and considering
the effect of making $\partial_z \phi \to \partial_z \phi + \lambda $ on
the
energy. A detailed derivation can be found in appendix
\ref{elasticity_to_bosons}.
In the vortex language, the expression  (\ref{density_op_bosonized})
has been derived directly in Ref. \onlinecite{giamarchi_vortex_long}
starting from the elastic theory of the  one dimensional vortex
lattice. However, in the case of a two plane system, the mapping on
bosons simplifies the description of vortex tunneling.

\subsection{Phase diagram of the vortex system}

\subsubsection{Flux lines in the absence of
disorder:}\label{twoplanes-nodisorder}

In the absence of disorder, the vortex system is equivalent to the
bosonic system in the absence of random potential discussed in section
\ref{boson_no_disorder}.
 The parameter $K_s$ controlling the stability of the
different phases depends on temperature, vortex density and elastic
constants of type vortex lattice as shown in
Appendix~\ref{elasticity_to_bosons}. Its expression is:
\begin{equation}
K_s = \frac{\pi \rho_0^2T}{(c_{44}c_1)^{1/2}}
\end{equation}

 Each of the four bosonic phases
translates into a phase of the vortex system.
In terms of flux lines, the CDW- is a crystal phase with a periodic
array of flux lines in each plane. The periodic array of vortices in the
second plane is translated of half a period along the x axis with
respect to the one that lies in the first plane in order to minimize
the repulsion energy. In that phase, there is \emph{no} interplane
hopping of
vortices. The CDW- is a strict 1d analog of a vortex lattice.

By contrast, the CDW$^{4\pi\rho_0}$ has the same disposition of vortices
but
they are constantly hopping from one plane to the other leading to an
effective period of the vortex lattice that is half the one of the
CDW-. This is a vortex lattice with ``superkinks'' from one plane to
the other. Due to the hopping between the planes this phase
is the equivalent in this restricted geometry of an entangled solid
phase.

The SF+ is a vortex liquid phase in which a flux line is making
``superkinks'' from
one plane to the other.
In this phase, vortices are unpinned. This implies strong dissipation
in the original vortex problem (i.e. in the vortex system
superconductivity is lost). Since in this phase hopping is relevant
vortex entanglement is expected. In other words, the SF+ phase maps
onto an entangled liquid.

On the other hand, the BPSF phase is a vortex liquid in which  flux
lines that are in different planes are bound to each other. No hopping
between planes takes place.

As usual in a two dimensional classical system, these phases
do not have true long range order but algebraically divergent
correlations. It is generally expected that by coupling weakly a lot of
two chains systems, the phase that would exist would be the one with
the most divergent fluctuations in the
isolated two chain system as suggested
by mean field theory. In the case of a coupled chain system, there
would be true superfluidity of the boson system (leading to an
exponential decay of density
fluctuations i.e. a true vortex liquid) or to formation of a vortex
lattice with true long range order.

In the case: $V(y)=-V_0\delta(y)$ and
$U({\mathbf r})=U_0\delta({\mathbf r})$,  $t_\perp$ is much larger
than $V$,
so that $K_s\simeq K_a\simeq K$ . This case
corresponds to the dotted line on Fig. \ref{fig:pure_bosonic}.

\begin{figure}
 \centerline{\epsfig{file=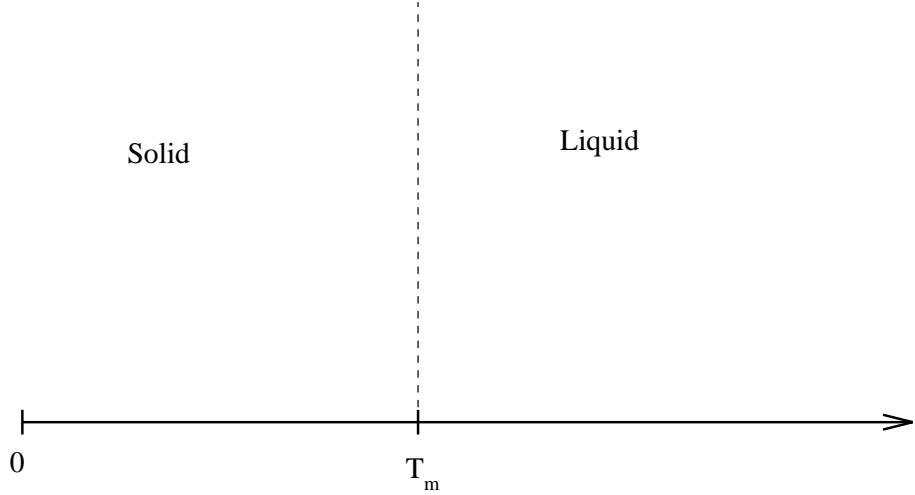,angle=-90,width=12cm}}
   \vspace{0.5cm}
\caption{The phase diagram of the pure two coupled planes
vortex system. For $T<T_m$ one has the solid phase, with vortices
confined in the planes, and for $T>T_m$, one has the liquid phase with
vortices hopping from plane to plane.}
\label{fig:pure-phases-t}
\end{figure}

As a consequence,  this system has 2 phases: a CDW- for $K_s< 1/2$, and
the SF+
phase for $K_m>1/2$. We assume here $V \sim t_\perp$ for simplicity
otherwise the boundary is displaced according to
Figure~\ref{fig:phasediag}.
The behavior of vortices in these
two phases is represented on figure \ref{fig:phases}.
In terms of vortices, there is a critical
temperature (see appendix \ref{elasticity_to_bosons})
$T_m=\frac{(c_{44}c_1)^{1/2}}{2\pi\rho_0^2}$ such that for $T<T_m$ the
vortex system has strong fluctuations towards the vortex lattice
state, and for $T>T_m$ strong fluctuations towards the vortex liquid
state as shown on figure \ref{fig:pure-phases-t}.

When vortex density (i.e. applied magnetic field) is increased, the
``melting temperature'' $T_m$ decreases. Also, the larger the elastic
constants, the higher the melting temperature. Such results are very
similar to the higher dimensional phase diagram.

Note that since the $\phi$ and $\theta$ fields are conjugated
an intermediate
``supersolid'' phase where one would have both a crystalline and a
superfluid order, cannot exist in this system,
even if perturbatively both operators would be
relevant for $1/4 < K < 1$. A more detailed proof is
given for the special temperature corresponding to $K=1/2$ in
Appendix~\ref{ap:super}. Whether such supersolid phase can exist for
an infinite number of planes as advocated in
ref.~\onlinecite{balents_smectic} remains  an open question.

\subsubsection{Effect of columnar disorder on the vortex system}

In the presence of columnar disorder, the vortex system at finite
temperature is equivalent
to the two chain boson system in the presence of a random potential
discussed in section \ref{boson_localization}.

If the interplane interactions are not too strong $K_s \sim K$ and
the various phases can be read from figure~\ref{fig:random_bosonic}.
The pinning of vortices is then controlled by the parameter
 $K_s=\frac{T}{2T_m}$, where $T_m$ is the melting temperature.

\begin{figure}
 \centerline{\epsfig{file=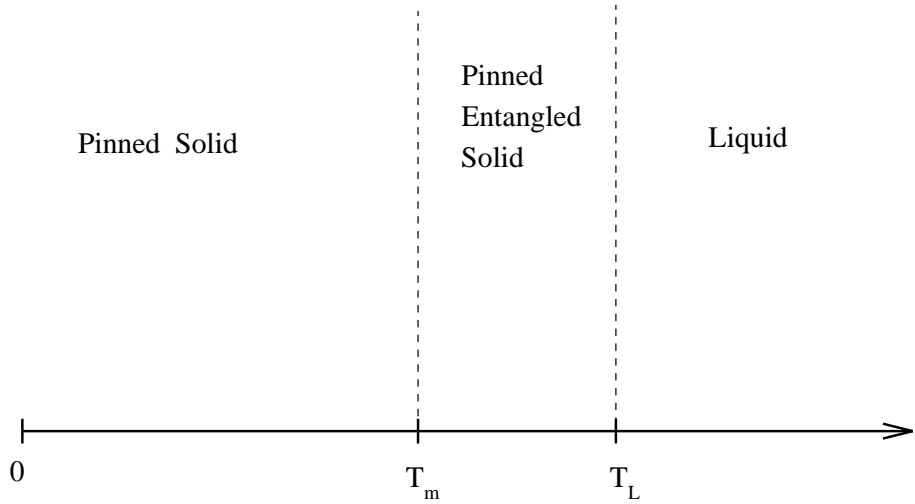,angle=-90,width=12cm}}
   \vspace{0.5cm}
\caption{The phase diagram of the weakly disordered two coupled planes
vortex system. For $T<T_m$, one has the pinned solid phase, for
$T_m<T<T_L=\frac 3 2 T_m$, one has the entangled pinned solid, in which vortex
pinning is weaker and vortices can hop from plane to plane. For
$T>T_L$, one goes into a vortex liquid phase in which vortices are
completely unpinned.}
\label{fig:phases-t}
\end{figure}

The main effect is to replace the solid-liquid transition that existed
in the pure system at  $T=T_m$ by a double transition.
The solid phase is replaced by a pined solid corresponding
to the $2\pi\rho_0$vortex lattice phase. This phase is a Bose glass.
At the pure melting temperature $T_m$, the solid phase disappears
and interplane interactions become irrelevant. The Bose glass
phase is replaced by a pinned CDW$^{4\pi\rho_0}$  vortex glass
phase. Since this phase
has dominant symmetric density fluctuations is possesses some
(disordered) solid type order. On the other hand interplane hopping
being relevant, this phase could also be viewed as a pinned solid with
entanglement, with intermediate properties between a solid and a
liquid.
It is only above a higher temperature $T_L = 3/2 T_m$ that this
entangled solid disappear to give place to the genuine liquid,
where disorder is irrelevant, as shown on figure \ref{fig:phases-t}. 
As is expected in the presence of columnar disorder
\cite{larkin_melting} the melting temperature is renormalized upward.

In the two plane vortex  system, the pinning of the $2\pi\rho_0$ vortex
lattice is much stronger
than the pinning of the $4\pi\rho_0$ vortex lattice, as can be shown
by the calculation of the pinning lengths and critical currents
detailed in the following section.
The reason is that the $4\pi\rho_0$ vortex lattice
is  destabilized by formation of superkinks that permit to
average the disorder and reduce its efficiency in pinning the vortex
system. This superkink formation is possible only when temperature is
large
enough to overcome the deformation energy that is needed to bend the
vortices, i.e. $T>T_m$.
In the $2\pi\rho_0$ vortex lattice
on the other hand, there is an enhancement of pinning due
to strong vortex repulsion
that  makes the vortex system
much more sensitive to the antisymmetric component of the random
potential.
For finite disorder since the pinning of the $2\pi\rho_0$ vortex
lattice is much stronger than
the one of the $4\pi\rho_0$ vortex lattice the transition line will be
shifted
upward from $K_a = 1/2$, i.e. to temperatures higher than $T_m$.
Columnar defects obviously tend to make hopping between
planes less relevant. However, we are in the limit where hopping
dominates over columnar disorder, so that the difference in columnar
pinning
potential between the planes is simply averaged out.
Such effect  appears nevertheless in renormalization group treatment around
the decoupled chains fixed point.

It is interesting to note that if the disorder is perfectly
correlated between the planes, as for example would be the case for
twin boundaries, it
can only couple to the $4\pi\rho_0$ component of the density in both
the $2\pi\rho_0$ vortex lattice and the $4\pi\rho_0$
vortex lattice, leading to the same pinning lengths in the two phases
up to prefactors.
This is easily seen by putting $V_1=V_2$ in Eq.
(\ref{boson_boson_random}).
Such result is due to the fact that the disorder
being the same in the two planes, it does not couple to the
antisymmetric density of the vortex solid phase, and
the averaging effect of vortex
hopping cannot lead any more to a reduction of vortex pinning.

As we have seen above, the smaller $\frac T {T_m}$ the stronger the
pinning.
Thus, at a given temperature,
a stiff vortex lattice having a higher $T_m$
 is much more easily pinned than a softer one.
Also, by  reducing vortex density (i.e.  reducing the
applied magnetic field) one increases $T_m$ and thus
 the pinning of the vortex lattice.

\subsection{Physical discussion}
\subsubsection{Weak disorder limit}

The two planes vortex system shows
 two superconducting phases with different vortex pinning lengths.
The first one, the $2\pi\rho_0$ vortex solid,
 is dominated by interplane vortex
repulsion and has the
 shortest vortex pinning
length : $l_{2\pi \rho_0}\sim \left(\frac 1 D\right)^{\frac 1
{3-K_s}}$ (see Eq. (\ref{loc_length_1chain})).

Expressed in terms of temperature, this gives $l_{2\pi \rho_0}\sim
\left(\frac 1 D
\right)^{\frac {T_1} {3(T_1-T)}}$, where $T_1=6T_m$.
 It is a Bose Glass phase, and at the
 melting point, the vortex pinning length becomes infinite with an
 essential singularity. Such singularity is typical of a two
 dimensional system.
  The other superconducting phase, the $2\pi\rho_0$ phase is dominated
by
formation of superkinks and has a larger
pinning length given by Eq. (\ref{pin_length_bos-hop})
$l_{4\pi\rho_0}\sim \left (\frac 1 {\overline{D}} \right)^{\frac 1
{3-4K_s}}$
 since the pinning potential is averaged out on the planes.
In terms of temperature, the behavior of the pinning length in this
second phase is $l_{4\pi\rho_0}\sim \left( \frac 1 {\overline{D}}
\right)^{\frac
{T_L}{3(T_L-T)}}$, where $T_L=\frac 3 2 T_m$.
 It is again a Bose Glass phase, but the underlying
lattice has half the period of the underlying lattice of the previous
phase. Similarly to the previous phase, the correlation length shows
an essential singularity at the melting point.
These two phases could be probed in neutron scattering experiments,
and could be distinguished by measuring the wavevector of the
underlying lattice.

A quantity of interest for vortices is of course the critical current.
A proper calculation of the critical current involves a
full solution
of the dynamics of vortices whereas all our
 calculations are restricted to thermal
equilibrium. Nevertheless, a simple estimate of the critical current
along the lines of the Larkin Ovchinnikov argument can
be made using only the equilibrium results.
In the Larkin-Ovchinnikov theory,
$R_c$  designates the distance beyond which the
random force approximation breaks down. It is argued that for
$r\gg R_c$ the vortices are pinned as independent bundles.
In the Larkin-Ovchinnikov theory, the balance between random force due
to disorder and the force $F_c$ due to the critical current implies
that :
\begin{equation}
F_c=\left(\frac{D} {R_c}\right)^{\frac 1 2 }
\end{equation}
for  vortices confined to planes parallel to the magnetic field in the
presence of columnar disorder.

The distance beyond which vortices are uncorrelated is just
the localization length of the bosons.
Replacing $R_c$ with the boson localization length gives a power law
dependence of the critical force on disorder.
It can be checked first that in the limit $T\to 0 $ one recovers the
Larkin-Ovchinnikov
result and second that the critical current goes to zero at the
localization-delocalization
transition for the bosons indicating the onset of dissipation in the
vortex problem.
In the pinned $2\pi\rho_0$ vortex solid phase the critical
current is given by:
\begin{equation}
F_c\sim D^{\frac {8T_m-T}{2(6T_m-T)}}
\end{equation}
In the pinned $4\pi\rho_0$ entangled vortex solid phase one has:
\begin{equation}
F_c \sim {\overline{D}}^{\frac{2T_m-T}{3T_m-2T}}
\end{equation}

\begin{figure}
 \centerline{\epsfig{file=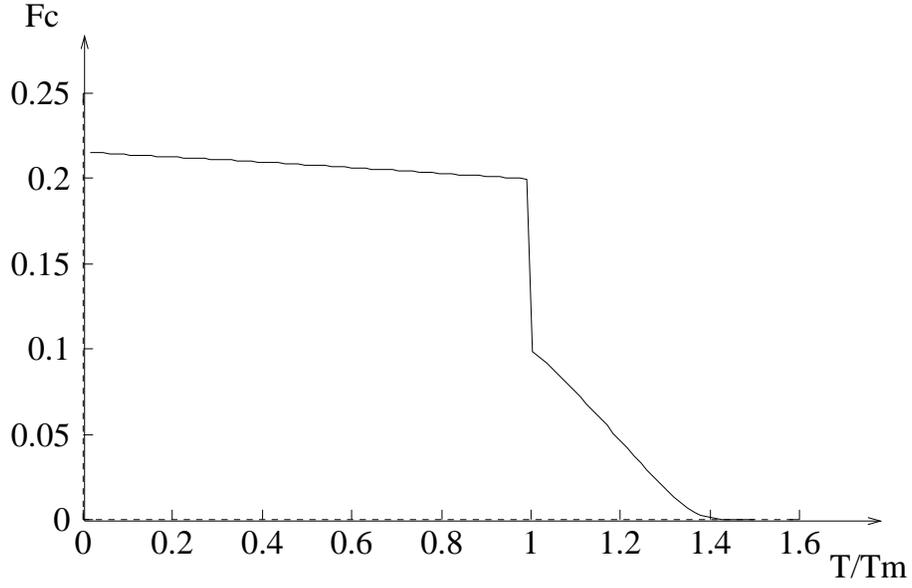,angle=-90,width=12cm}}
   \vspace{0.5cm}
\caption{Dependence of critical current upon temperature
for $D=0.1$. For $T>T_m$, entanglement appears and the critical
current is strongly reduced. For $T>\frac 3 2 T_m$, the entangled
solid is melt leading to a zero critical current. }
\label{fig:fc-t}
\end{figure}

The $4\pi\rho_0$
Bose
Glass does not extend down to 0K, so that there is no reason
to expect recovering the Larkin-Ovchinnikov result for $T\to 0$.
 For $T<T_m$ it is replaced by the
$2\pi\rho_0$ Bose Glass as explained in section
\ref{1plane_vs_2planes}.
At $T=T_m$, there is a transition from the $2\pi\rho_0$ vortex liquid
to the $4\pi\rho_0$ vortex liquid.
At that point(see figure \ref{fig:fc-t}), the critical current drops
from $F_c\sim D^{7/10}$ for
$T\to T_m^-$ to $F_c\sim D$ for $T\to T_m^+$.
This apparent discontinuity is a consequence of the approximations
made.
However, one expects a very rapid drop of $F_c$ as $T$ traverses
$T_m$.
\subsubsection{Qualitative discussion of the strongly disordered case}

\begin{figure}
 \centerline{\epsfig{file=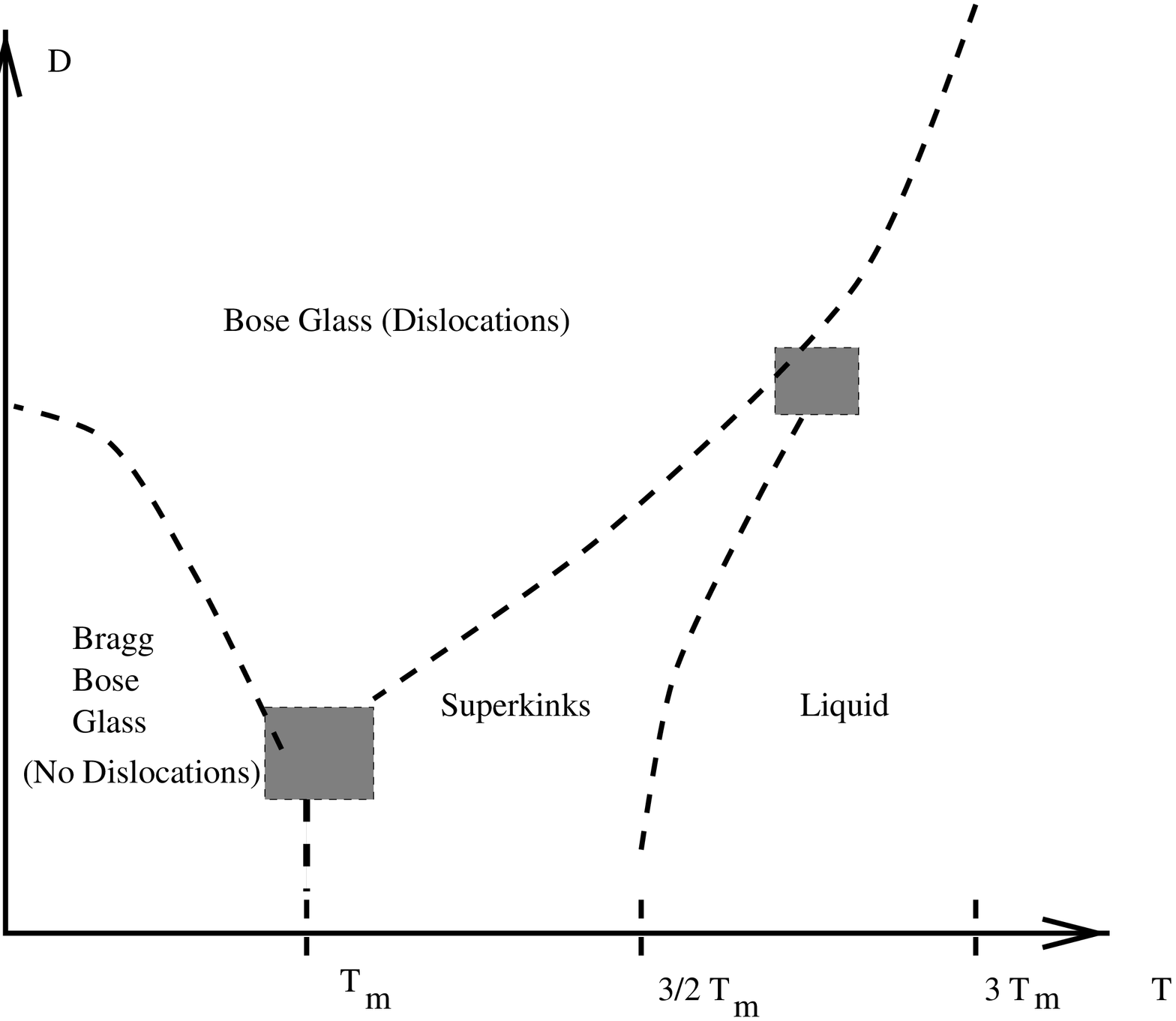,angle=-90,width=12cm}}
   \vspace{0.5cm}
\caption{Phase diagram as a function of temperature and disorder.
For large disorder, a Bose glass with dislocations is formed. At weak
disorder, depending on temperature, there is a Bragg Bose glass
(without dislocations), an entangled Bose Glass and a vortex
liquid. The shaded squares correspond to regions where a multicritical
point may lie. }
\label{fig:strongdisorder}
\end{figure}

In the strongly disordered case, disorder can localize bosons before
interchain hopping or interchain repulsion can open a gap in the
antisymmetric degrees of freedom. In the case of vortices, this
corresponds to independent pinning of vortices in each plane with
neither  interplane repulsion of vortices nor hopping of vortices.
In such case,the two planes are completely decoupled and
there is no correlation of vortex positions between the
two planes. This phase can be seen as a pinned solid containing
dislocations between the planes.
This transition is the analogous of the transition between a
Bragg glass phase and a vortex glass one occurring for point
like disorder \cite{carpentier_bglass_layered,kierfeld_bglass_layered}.
In the case of correlated disorder it corresponds to the transition
between a dislocation free Bose glass
(or Bragg Bose glass \cite{giamarchi_diagphas_prb})
to the ``standard'' Bose glass containing dislocations.
The condition for existence of that phase is that the pinning length
of vortices in one plane is much shorter than both the correlation
length due to interplane repulsion and the correlation length due to
interplane hopping.
This leads to:
\begin{eqnarray}\label{eq:condition_for_dislocated_phase}
  D \gg V^{\frac{3T_m -T}{2T_m-T}} \\
  D \gg {t_\perp}^{\frac {T(3T_m-T)}{2T-T_m}}
\end{eqnarray}
The resulting phase diagram is drawn on figure
\ref{fig:strongdisorder}.
The transition line between the phase with ``superkinks'' and the Bose
 Glass without dislocations corresponds to $ V^{\frac{3T_m
    -T}{2T_m-T}}={t_\perp}^{\frac {T(3T_m-T)}{2T-T_m}}$. This vertical line
ends at a point $D={t_\perp}^{\frac {T(3T_m-T)}{2T-T_m}}=V^{\frac{3T_m
    -T}{2T_m-T}}$. This suggests a possible multicritical point
separating a Bose glass phase with dislocations, a Bose glass phase
without dislocations (i.e. a Bragg Bose Glass) and a Bose Glass phase
with ``superkinks''.

For $T>3T_m$, the Bose glass phase with dislocation disappears and is
replaced by a liquid phase.
The transition line between the pinned entangled solid and the liquid
phase starts at $T=\frac 3 2 T_m$ for $D \to 0$. The melting
temperature of the pinned entangled solid increases with disorder
since the pinned solid gains energy from the random potential
with respect to the liquid.
The melting line of the Bose glass with dislocations can cut the
melting line of the Bose glass with ``superkinks''. The intersection
may correspond to a second multicritical point. The two possible
multicritical points lie in the shaded zones of Fig. \ref{fig:strongdisorder}.
It is possible to obtain the critical currents by the method of the
previous section. One obtains:
\begin{equation}
F_c=D^{\frac{4T_m-T}{6T_m-2T}}
\end{equation}

\subsubsection{comparison of the one plane and the two plane vortex
system}
\label{1plane_vs_2planes}
For the two planes vortex system, pinning depends on two factors. The
first one is of course temperature. At high enough temperatures,
vortices always depin.
The second factor is the relative magnitude of $t_\perp$ and $V$.
$t_\perp$ depends on the potential that confines vortices in the
planes
but also on the temperature via the unperturbed wavefunction (see
appendix for the case of a $\delta$ potential).
$V$ depends both on temperature and vortex-vortex repulsion. It also
depends on the potential that confines vortices in the plane via the
unperturbed
 wavefunction.
Therefore, the phase diagram of the non-disordered two planes system
as a function of temperature is strongly affected by the potential
that are chosen to model vortex-vortex interactions and vortex planes
interactions.
It is  known from section
\ref{twoplanes-nodisorder}
that in the absence of disorder the two plane vortex
system is in a vortex lattice state for $T<T_m$ and in a vortex liquid
 state for
$T>T_m$. Also, from section \ref{twoplanes-nodisorder} we know that
 for $T<T_m$, we have a
$2\pi\rho_0$ Bose Glass, for $T_m<T<T_L$, a pinned $4\pi\rho_0$
vortex
lattice containing superkinks, and for $T>T_L$ vortices get unpinned.

Remarkably, the two temperature are related by the simple relation
$T_L=\frac 3 2 T_m$. It is also interesting to remark that at $T=T_m$
there is a discontinuity in vortex pinning length and in critical
current. Of course, such a discontinuity is the result of the neglect
of disorder with respect to the gap, and a more exact treatment would
lead to a continuous albeit extremely rapid decrease of the critical
current with temperature.

For the one plane system on the other hand,
there is only one Bose glass phase and
vortex depinning is solely driven by temperature.
The transition Bose Glass to vortex liquid occurs at a higher
temperature $T_{\text{1 plane}}=2T_L$ and for given temperature and
given amount of disorder the vortex pinning length is shorter in the one
plane system if $T>T_m$. On the other hand, for $T<T_m$, the pinning
length of the two planes system is the shortest due to the transition
from the $4\pi\rho_0$ to the $2\pi\rho_0$ Bose glass as the temperature
is
lowered.
As a consequence, for $T<T_m$ vortex pinning is the strongest in two
planes systems whereas for $T>T_m$ it is the strongest in one plane
systems.
What distinguishes sharply one plane vortex systems from two planes
vortex systems is the existence of a two step transition from the Bose
glass phase to the vortex liquid. The first step is the formation of
superkinks that halve the periodicity of the underlying lattice,
leading to a second Bose glass phase much less well pinned by columnar
disorder, and in a second step the melting of the new Bose glass
phase.

All these considerations are valid for $T>0$. However, for $T=0$,
thermal fluctuations are not available to screen the pinning
potential. It is therefore interesting when comparing the one plane
and the two plane system to consider the limit $T\to 0$.
In the $T\to 0$ limit, it is possible to obtain the pinning length by
an Imry-Ma \cite{fukuyama_pinning} type argument.
One requires that the phase of the field coupled to the random
potential jumps by  $2\pi$ on a length $L$.
This leads to a variation of $\phi_s$ on the length $L$ of $\delta
\phi_s \pi \sqrt{2}$.
The resulting pinning energy is:
\begin{equation}
E(L)= \frac{2\pi^2c_1}{L^2}-\sqrt{\frac{2D}{L}}
\end{equation}
Minimizing with respect to $L$ gives:
\begin{equation}
L_{\text{pin.,2ch.}}=\left( \frac {32 \pi^4 c_1^2} D\right)^{1/3}
\end{equation}
For one plane, the same procedure gives :
\begin{equation}
L_{\text{pin., 1 ch.}}=\left( \frac {16 \pi^4 c_1^2} D\right)^{1/3}
\end{equation}
The pinning length is therefore larger in the two plane system by a
factor $2^{1/3}$ than in the one chain system.

Therefore, at T=0, we expect two regimes:
For $D^{1/3} \gg V^{1/2}$, the two planes are mutually decoupled and
individually pinned on disorder. For $D^{1/3} \ll V^{1/2} $, the two
planes are locked and the pinning length is \emph{larger} than in a
single plane system with the same disorder strength.
This situation is reminiscent of the one that obtains in higher
dimensions with a solid phase without dislocation at weak disorder and
a more strongly pinned solid phase with dislocation at higher
disorder. Such situation is expected on physical grounds, since
forming dislocations allows the system to gain more energy from the
random pinning potential.

At non-zero temperature, the solid phase with dislocations is more
easily melt than the solid without dislocations. Therefore, one
expects that the pinning length decreases faster with temperature in
the phase in which the vortices are pinned independently in the planes
than in the phase in which the vortices are pinned by the
antisymmetric component of disorder. Such behavior is indeed what we
have obtained for $T>0$.

\section{Conclusion}  \label{sec:conc}

In this paper, we have investigated two chains of strongly correlated
bosons, coupled by boson interchain hopping and boson-boson repulsion,
and applied our results both to quantum systems and to their classical
analog of vortices in type II superconductors in the presence of
columnar defects.

Using bosonization and renormalization techniques,
we have obtained the phase diagram in the absence of disorder. It
contains four different phases: two charge
density wave phases of different periodicity, a conventional superfluid
phase,
and a superfluid phase made of pairs of bosons. This last superfluid
phase is an analog of the an XY phase found in spin one
\cite{schulz_spins}
chains and spin ladders\cite{orignac_2spinchains}. For weak inter chain
interactions one finds a transition between a superfluid and
``crystalline'' (charge density wave) phase. This transition occurs
for considerably more repulsive interactions ($K = 1/2$) than in the
case of a single chain, where it would occur for hard core bosons
($K=1$). Interchain hopping thus stabilizes considerably the superfluid
phase. Paradoxically, the long wavelength part of the interchain
repulsion favors hopping between the chains and thus also helps in
stabilizing a superconducting phase.

We have then considered the
effects of a random potential. For a single chain this leads to the
localization of the bosons for moderately repulsive interactions
($K=3/2$). In the case of the ladder the superfluid phase is much
more resistant to localization. The transition occurs at more repulsive
interactions ($K=3/4$). In particular bosons with only contact
interactions are now always superfluid. Two different
localized phases exist: first a pinned charge density wave where
interchain hopping occurs, then for more repulsive interactions ($K<
1/2$), a pinned charge density wave where hopping is frozen and with
``crystalline'' order of the bosons. Interchain boson repulsion pushes
the system towards localization. Using renormalization we have also
extracted the localization length as a function of disorder strength and
interactions, and the temperature and frequency dependence of the
conductivity. For the ladder, the exponent of the superfluid correlation
function takes at the transition the universal value $\eta = 1/3$.
The results for the compressibility and exponents obtained in
our microscopic model are in agreement with the general scaling
relations proposed for the superfluid-insulator
transition\cite{fisher_bosons_scaling}.
Moreover, the superfluid correlation exponent $\eta$ for the ladder
takes the same value than the one for a single chain\cite{giamarchi_loc}, suggesting the
interesting possibility of a \emph{universal} exponent of the
superfluid-insulator transition in one dimension. Such a property would
be much stronger than the simple inequality $\eta < 1$ suggested by the
scaling. For an even number of chains, we have given
a general argument, valid when all the mode but one are gapped, that
this is indeed the case. This conjecture remains of course to be tested
in much more details, both for the case of an odd number of chains and
for situations where some modes could remain gapless.

These results have a direct translation for the case of vortices in type
II superconductors. The bosonic ladder corresponds here to two planes of
vortices with both vortex-vortex interactions and hopping of vortices
between the planes. In the boson vortex mapping superfluidity of the
boson corresponds to a liquid state for the vortices, whereas the CDW
phases correspond to crystalline order of the vortices.
For the pure systems there is a transition between a high temperature
liquid phase and a low temperature solid phase. The melting temperature
$T_m$ is much smaller than the melting of a single plane due to the
possibility
of forming superkink between the planes. The transition is direct
without an intermediate ``supersolid'' phase having both crystalline and
superfluid properties, advocated to exist in this
geometry\cite{balents_smectic}. This absence
of supersolid phase could be due to our restriction to only two coupled
planes instead of an infinite number. However since the existence of the
supersolid was only advocated based on perturbative arguments that would
also give such a phase for the ladder, our study of the ladder shows
that the possible existence of such a phase needs to be put on a firmer
basis. In addition to these two
phases, stable for weak interplane interaction, two other phases can
exist: a vortex solid but where the vortices form superkinks between
planes and a bound vortex pairs liquid.
The random potential in the bosonic system is equivalent to columnar
defects for the vortices. In presence of disorder two different
transitions take place. At low temperature one is in a pinned solid
state with no hopping between the planes. This corresponds to a Bose
glass state for the vortices without dislocations between planes
(nicknamed Bragg Bose glass\cite{giamarchi_diagphas_prb}). Upon increasing the temperature a first
transitions occurs at $T_m$ between the Bose glass and a phase where
vortices hop between planes giving a (pinned) solid with half the
periodicity of the low temperature one. Combining RG with
Larkin-Ovchinnikov arguments allows to compute the critical currents.
At this transition  the critical current is strongly decreased but
remains finite. This decrease of the critical current is typical of
columnar disorder and would not occur for pinning by twins.
The liquid, with zero critical current is only recovered
at higher temperature $T = 3/2 T_m$ both for twins and columnar
defects.
This increase of the melting
temperature is reminiscent of what is expected for columnar defects in
higher dimensionality.
Note that in our one dimensional system even an infinitesimal disorder
would cause a finite shift of the melting temperature.

If disorder is increased, the planes decouple or equivalently
dislocations appear between them. One recovers in this case the
properties of individually pinned planes. This leads to an increase of
the critical current similar to the one occurring for point like
disorder when going from a Bragg glass state to a vortex glass state\cite{giamarchi_diagphas_prb}.
The transition occurs here between a dislocation free Bose glass (Bragg
Bose glass) and a conventional Bose glass.
Upon increasing the temperature there is then a single melting to a
liquid phase but at a much higher temperature $T' = 3 T_m$.
One thus expects for correlated disorder the same topology of phase
diagram as for point like disorder, with two different glass phases
 but with the difference that  increasing the correlated disorder
raises the melting temperature. Whether the superkink phase close to the
liquid for weak disorder exists in a three dimensional system is an
interesting question that would deserve more investigations.

The ladder geometry thus provide an interesting starting point to study
competition between elasticity and disorder in vortex systems while
retaining realistic feature. It would  be interesting to investigate
the effect of point disorder
and the interplay of columnar and point disorder   in that
system. This has already been done for a single plane system
\cite{hwa_compet_disorder} and it would be interesting to know how
interplane hopping modifies the response of the system to competing
disorder.
Also, systems with
more than two coupled planes could provide further insight in the
properties of coupled vortex planes and may be relevant to
experiments. Finally, a mean field theory investigation of coupled
planes would help in comparing the results with experiments on vortex
pinning in BiSSCO.

\acknowledgments

We are grateful to D. Carpentier, P. Le Doussal and H.J. Schulz,  .
for many illuminating discussions.

\appendix
\section{Bosonization of bosons} \label{reminder}

In this section the semi-phenomenological bosonization procedure
introduced in Ref. \onlinecite{haldane_bosons} to describe the low
energy physics of one dimensional
interacting bosons is reviewed briefly. The reader already familiar
with this technique should read it quickly to get acquainted with the
notations.
In Ref. \onlinecite{haldane_bosons} it  is argued that a one dimensional
 interacting boson system always has an associated
 bosonized Hamiltonian describing the low energy, long momentum
phonon excitation of the system ~:

\begin{equation}
\label{boson_formulas}
H=\int \frac{dx}{2\pi}\left[ uK(\pi\Pi)^2 + \frac{u}{K}(\partial_x
\phi)^2 \right]
\end{equation}
$\Pi$ and $\phi$ satisfy canonical commutation relations~:
$[\Pi(x),\phi(x')]=\imath \delta(x-x')$.
According to ref.~\onlinecite{haldane_bosons}, the operators of the
original bosonic theory can be expressed as a function
of the fields entering the Hamiltonian (\ref{boson_formulas}).
The bosonized form of the boson annihilation operator in one dimension
is~:
\begin{equation}\label{annihilation_op_bosonized}
\psi_{B}(x)\sim e^{\imath \theta(x)}\left(
\sum_{m=-\infty}^\infty e^{2\imath m(\phi(x)+\pi\rho_0 x)} \right)
\end{equation}
where $\theta(x)=\pi \int_{-\infty}^x dx' \Pi(x')$ and $a$ is a short
distance cutoff.

The bosonized form for the density is~:
\begin{equation}\label{density_op_bosonized}
\rho_{B}(x)=(\rho_0 +\frac{\partial_x
\phi}{\pi})\sum_{m=-\infty}^\infty e^{2\imath m(\phi(x)+\pi \rho_0 x)}
\end{equation}
where $\rho_0$ is the average boson density.
Boson Green's functions
and density density correlation functions
are obtained by the standard methods
of bosonization \cite{haldane_bosons,solyom_revue_1d,emery_revue_1d}
giving:

\begin{eqnarray} \label{correlations}
\langle T_\tau e^{\imath n\theta(x,\tau)}e^{-\imath n\theta(0,0)}\rangle
=\left(\frac 1 {x^2+u^2\tau^2}\right)^{\frac {n^2} {4K}} \nonumber \\
\langle T_\tau e^{\imath 2n\phi(x,\tau)}e^{-\imath 2n\phi(0,0)}\rangle
=\left(\frac 1 {x^2+u^2\tau^2}\right)^{n^2 K} \nonumber \\
\end{eqnarray}
The boson creation operator is also the order parameter for
superfluidity and the $2\pi\rho_0$ component of the density is the
charge
density wave order parameter.
Thus,  (\ref{correlations}) shows that in the one dimensional bosonic
system there is no long range order (in agreement with the
Mermin-Wagner theorem \cite{mermin_wagner_theorem,mermin_theorem}), but
only
quasi long range
order. In particular, the susceptibilities associated with these
operators diverge in the long wavelength low frequency limit.
When $K$ increases, the system has a stronger tendency to superfluidity,
whereas
when $K$ decreases it has a stronger tendency to charge density wave
order.

 $u,K$  still have to be determined
as a function of the microscopic boson-boson interactions.
To do so,  one has to
relate them to ground state parameters of the interacting system.
In Galilean invariant systems
, an exact relation between $u,K$ holds\cite{haldane_bosons} and
simplifies
considerably the situation.
This  relation is : $uK=\frac{\pi\rho_0}{m} $
, where $m$ is the mass of a boson.
Knowing this, only one relation between $u,K$ and a parameter of the
ground state suffices to determine $u$ and $K$.
 A convenient quantity is the compressibility $\chi$  defined by:
\begin{equation}
\label{compressibility-def}
\chi=\frac1L \left(\frac{\partial^2 E}{\partial^2 L}\right)^{-1} \\
\end{equation}
 E being the ground state energy of
the system.
The compressibility can also be expressed as a function of $u,K$ in the
form:
\begin{equation}
\label{compressibility}
\chi=\frac{K}{\pi u\rho_0^2}
\end{equation}
and permits a determination of $u,K$ as a functions of the parameters
of the microscopic Hamiltonian. $\chi$ contains all interaction
effects for Galilean invariant systems. On physical ground, the higher
the boson-boson repulsion, the lower the compressibility.
Therefore, $K$ decreases when  repulsion increases. According to
 (\ref{correlations}) this means that repulsive interactions tend to
favor
charge density wave order versus superfluidity.
In the case of bosons interacting with a repulsive
$\delta$-function potential, the ground state energy and the
compressibility can be obtained
exactly as a function of the interaction \cite{lieb_bosons_1}. This
allows for the determination
of $u,K$. In the case of a more general interaction potential, one has
to rely on numerical
techniques such as Quantum Monte Carlo simulations
\cite{batrouni_bosons_numerique}.

Bosons on a lattice lack Galilean invariance, so that $uK \neq
 \frac{\pi\rho_0}{m}$. Nevertheless,  (\ref{compressibility}) remains
valid.
 To determine
$u,K$ one needs another relation linking $u,K$ to a parameter of
the ground state.
One convenient parameter is the charge stiffness \cite{kohn_stiffness}
 defined for a closed ring of perimeter $L$ by:
\begin{equation}
\label{stiffness-def}
{\mathcal D} =\frac{L}{2}
\left(\frac{d^2 E(\varphi)}{d^2 \varphi^2}\right)_{\varphi=0}
\end{equation}
 $\varphi$ being
 a flux threading the system.
 For bosons that do not carry electrical charge, it is nevertheless
possible
 to define a "charge stiffness" by using the sensitivity of energy
levels
 to boundary conditions since the only effect of the flux $\varphi$
amounts to
 a change of boundary conditions.
 From the definition  (\ref{stiffness-def}), one obtains~:
\begin{equation}
\label{stiffness_bos}
{\mathcal D}=uK
\end{equation}
In Galilean invariant systems, the charge stiffness
is independent of the interactions.
Combined  with  (\ref{compressibility}) this equation  allows to
determine
$u,K$  as a function of the original parameters of the Hamiltonian in
the most general case.
For instance, for hard core bosons on a lattice, using an exact mapping
onto
the XXZ spin 1/2 chain in a magnetic field \cite{fisher_xxz} one can
obtain
an exact expression of $u,K$ using the Bethe Ansatz solution of the spin
chain problem.
The more general case of soft core boson needs to be treated
numerically.
\section{Superfluid-disordered exponent} \label{ap:simple}
Let us consider a ladder with $p$ legs. The density on chain $i$ can
be expressed in terms of the field $\phi_i$ as
\begin{equation}
\rho_i \sim e^{2 \imath \phi_i}
\end{equation}
As for the two leg ladder it will be convenient to introduce the various
modes (equivalent to the symmetric and antisymmetric ones) between the
chains.
 If one now assumes that every mode except the
totally symmetric one is gapped (the fact that the totally symmetric
one remains always gapless is a consequence of the Galilean invariance),
one can then proceed as for the two leg ladder. The totally symmetric
mode has the following expression:
\begin{equation}
\phi_s=\frac 1 {\sqrt{p}}\sum_{i=1}^p \phi_i
\end{equation}
And all the modes can be decomposed as:
\begin{equation}
\phi_i=\frac{\phi_s}{\sqrt{p}} + \text{non-symmetric modes}
\end{equation}
Let us assume that all of the non-symmetric modes are gapped with an
average value for the corresponding $\theta$'s fields.
The coupling to disorder
should only retain the massless mode. It is thus necessary to form
higher order terms in perturbation theory to cancel all the other modes.
It is easy to check that the remaining coupling to disorder is of the
form
\begin{equation} \label{perturb}
H = \int dx  V(x) e^{\imath 2 p/\sqrt{p} \phi_s{x}}
\end{equation}
since it is necessary to go to $p$-th order to get rid of all the other
modes. (\ref{perturb}) obviously reproduces the two leg result.
Using the same RG argument the transition now occurs for $K = 3/(2p)$,
i.e. more and more repulsive interactions.
Similarly the superfluid order parameter on chain $i$ is given by
\begin{equation}
\psi_i \sim e^{\imath \theta_i}
\end{equation}
Using again the decomposition in the above modes \emph{and} assuming
that all gaped modes have a an average value for the $\theta$ field, one
gets
\begin{equation}
\psi_i \sim e^{\frac{\imath} {\sqrt{p}} \theta_s} C
\end{equation}
where $C$ is a universal constant coming from the massive modes.
In that case the correlation function at the transition decays as
\begin{equation}
\langle \psi_i(r) \psi_i(0) \rangle \sim
\left(\frac{1}{r}\right)^{\frac{1}{2p K}}
\end{equation}
giving the universal exponent $\eta = 1/3$ regardless of the number of
legs.

\section{Derivation of the Hamiltonian of the coupled planes in the
tight-binding approximation}\label{tight-binding}
The Hamiltonian for non-interacting bosons in a plane with two
potential wells  is~:
\begin{equation}\label{two-wells}
H=-\frac {\hbar^2}{2m}(\nabla_y^2 +\nabla_x^2) + V(y-Y_1) + V(y-Y_2)
\end{equation}
$V(x)$ represents an isolated potential well.
 Consider the Hamiltonian h:
\begin{equation}\label{one-well}
h=-\frac {\hbar^2}{2m}\nabla_y^2 + V(y)
\end{equation}
and call $\varphi(y)$ its eigenfunctions.
In the tight binding approximation the eigenfunctions are sought in
the approximate form :
\begin{equation}
\Psi(x,y)=\psi(x) (\alpha\varphi(y-Y_1)+\beta\varphi(y-Y_2))
\end{equation}
and the Hamiltonian is of the form $h+V$.
The Hamiltonian is simplified into a Hamiltonian with a hopping term:
$t_\perp =\int dy \varphi(y-Y_1) V(y-Y_2) \varphi(y-Y_2)$
In the case~: $V(y)=-V_0 \delta(y)$, it is possible to obtain the
exact eigenfunctions of  (\ref{one-well}) and calculate $t_\perp$ as~:
\begin{equation}
t_\perp=-V_0\kappa \exp \left( -\kappa \mid
  Y_1-Y_2 \mid \right)
\end{equation}
with $\kappa=\frac {mV_0}{\hbar^2}$.
This allows a rewriting of the Hamiltonian in the second quantized
representation
as:
\begin{equation}
H=\int dx \left[\Psi_1^\dagger(x) -\frac
{\hbar^2}{2m}\nabla_x^2\Psi_1(x)
+ \Psi_2^\dagger(x) -\frac {\hbar^2}{2m}\nabla_x^2\Psi_2(x)
+t_\perp (\Psi_1^\dagger(x)\Psi_2(x) +\text{H. c.}) \right]
\end{equation}
The wavefunction (second-quantized) being:
$\Psi(x,y)=\Psi_1(x)\varphi(y-Y_1)+\Psi_2(x)\varphi(y-Y_2)$.
Adding interactions leads to the following expression for intrachain
interaction:
\begin{equation}
u(x-x')=\int dy dy'
U(x-x',y-y')\mid\varphi(y)\mid^2\mid\varphi(y')\mid^2
\end{equation}
and for interchain interaction:
\begin{equation}
v(x-x')=\int dy dy'
U(x-x',y-y')\mid\varphi(y-Y_1)\mid^2\mid\varphi(y'-Y_2)\mid^2
\end{equation}
There are also interactions which do not conserve the number of bosons
in one chain
However, these interactions merely correct the hopping term.
In the case of the delta-function potential well, and a delta
function repulsion between vortices in the form:
\begin{equation}
U(x-x',y-y')=U_0 \delta(x-x') \delta(y-y')
\end{equation}
 $u,v$ have the following expressions :
\begin{eqnarray}
  \label{interactions-for-2wells}
  u(x-x')=U_0\frac \kappa 8 \nonumber \\
  v(x-x')=U_0 \frac \kappa 8 \left[e^{-4 \kappa \mid Y_1-Y_2 \mid}+2
\kappa \mid Y_1-Y_2 \mid e^{-2 \kappa \mid Y_1-Y_2 \mid}\right]
\end{eqnarray}
The tight binding approximation is well justified in the limit $\kappa
\mid Z_1-Z_2 \mid \gg 1$ only.
Therefore in the case of delta functions interactions, interchain
hopping dominates over interchain repulsion.
This may change in the case of longer range interactions.

\section{derivation of the coupling with disorder for the two plane
system}
In the mapping of vortex planes onto coupled bosonic chains,
introduction of columnar disorder in the two plane system is
equivalent to
introduction of a time independent random potential in the two chain
system.
The part of the Hamiltonian describing such disorder is:
\begin{equation}\label{in-plane}
H_{\text{disorder}}=\int dx dy v(x,y) \Psi^\dagger(x,y) \Psi(x,y)
\end{equation}
with:
\begin{equation}
\overline{v(x,y)v(x',y')}=D\delta(x-x')\delta(y-y')
\end{equation}
y being the coordinate perpendicular to the chains and $\Psi$ being the
full second quantized wave function.
The chains being treated in the Tight Binding Approximation, one has:
\begin{equation}\label{tba}
\Psi(x,y)=\psi_1(x)\varphi(y-Y_1)+\psi_2(x)\varphi(y-Y_2)
\end{equation}
Porting  (\ref{tba}) into  (\ref{in-plane}) leads to the following
coupling:
\begin{equation}
H_{\text{disorder}}=\int dx \left( v_1(x) \psi^\dagger_1 \psi_1(x)
+v_2(x) \psi^\dagger_2 \psi_2(x)\right)
+\int dx \delta t_\perp(x)\left(\psi^\dagger_1 \psi_2(x) +
\psi^\dagger_2 \psi_1(x)\right)
\end{equation}
With:
\begin{eqnarray}
v_2(x)=\int dy v(x,y) \varphi^2(y-Y_2) \nonumber \\
v_1(x)=\int dy v(x,y) \varphi^2(y-Y_1) \nonumber \\
t_\perp(x)=\int dy v(x,y)\varphi^2(y-Y_2) \phi^2(y-Y_1)
\end{eqnarray}
These definitions lead to:
\begin{eqnarray}\label{averages}
\overline{v_1(x)v_1(x')}=\overline{v_2(x)v_2(x')}=D \delta(x-x') \int dy
\varphi^4(y) \nonumber \\
\overline{v_1(x)v_2(x')}=D \delta(x-x') \int dy
\varphi^2(y-Y_1)\varphi^2(y-Y_2) \nonumber \\
\overline{\delta t_\perp(x) \delta t_\perp(x')}=D \delta(x-x')
\varphi_1^2(y)\varphi_2^2(y)
\end{eqnarray}
If the potential that confines vortices in the plane is of the form
$-V_0(\delta(y-Y_1)+\delta(y-Y_2))$, one can obtain an explicit form of
the integrals in  (\ref{averages}).
Introducing $\kappa=\frac {mV_0}{\hbar^2}$, one has
$\varphi(y)=\sqrt{\kappa}e^{-\kappa y}$ and :
\begin{eqnarray}\label{explicit-averages}
\int^\infty_{-\infty} dy \varphi^4(y) =\frac \kappa 2 \nonumber \\
\int^\infty_{-\infty} dy  \varphi^2(y-Y_1) \varphi^2(y-Y_2)=
\kappa(1 + \frac { \kappa a} 2)e^{-2 \kappa a} \nonumber \\
\end{eqnarray}
with $a=\mid Z_1-Z_2\mid$ .
The tight binding approximation being valid only for $\kappa a \gg 1$,
$v_1$ and $v_2$ are almost uncorrelated.
Also, disorder induces a very weak random interchain hopping.
For the tight binding approximation to be valid, the random interchain
hopping has to be much smaller than $t_\perp$.
More precisely, the correlation length induced by the random hopping
terms has to be much larger than the correlation length induced by
$t_\perp$.
We know that : $t_\perp=-\kappa V_0
\exp \left( -\kappa \mid Z_1-Z_2 \mid \right)$.
This leads to the following criterion:
\begin{equation}
\left(\frac 1 {t_\perp}\right)^{\frac 2 {4-K_a}} \ll \left(\frac 1 D
\right)^{\frac 1 {3-K_a}}
\end{equation}
In terms of $V_0$ the criterion reads :
\begin{equation}
D \ll \frac{(\kappa V_0)^{\frac {2(3-K_a)}  {4-K_a} }}
{ \kappa (1 + \frac {\kappa a} 2) e^{\frac {2\kappa a} {(4-K_a)}}}
\end{equation}
Since $\kappa a \gg 1$ when the tight binding approximation is correct,
the
condition on $D$ is always satisfied as long as the tight binding
approximation
is correct.

\section{Relation between elastic constants of a one vortex plane system
and
the parameters of  its equivalent bosonic theory}
\label{elasticity_to_bosons}
We consider a plane containing vortices. Let $\rho_0$ the vortex
density, and $c_{44}$ and $c_1$ the elastic constants of the vortex
system.
The elastic Hamiltonian of the vortex system is:
\begin{equation}
H=\frac 1 2 \int dx dz \left[ c_{44} (\partial_z u)^2 + c_1
(\partial_x u)^2 \right]
\end{equation}
The Euclidean action of the equivalent bosonic system is:
\begin{equation}
S= \int \frac{dx d\tau}{2 \pi K} \left[ \frac{(\partial_\tau
\phi)^2} u + u (\partial_x \phi)^2\right]
\end{equation}
Identifying the path integrals, and reminding that $\phi(x)=\pi \rho_0
u(x)$
one finds:
\begin{eqnarray}
c_{44}=T\frac{\pi \rho_0^2}{uK} \nonumber\\
c_1=T\frac{\pi u \rho_0^2}{K}
\end{eqnarray}
 $c_1$ is the inverse compressibility \emph{both} of the
vortex system and the boson system. Also, for a translationally
invariant vortex system: $c_{44}=\rho_0 m$
Finally, $K=\frac {\pi \rho_0^2 T}{\sqrt{c_{44} c_1}}$

\section{Absence of a supersolid phase in the two plane vortex system}
\label{ap:super}
The discussion is done one the equivalent boson Hamiltonian.
At the point $K_a=1/2$, a refermionization of $H_a$ can be
performed\cite{shelton_spin_ladders}. 
One obtains:
\begin{equation}
H_a=\int \frac{dx}{2\pi} v_F\left[(\pi\Pi)^2+(\partial_x\phi)^2\right]
+\frac{t_\perp}{\pi \alpha} \int \cos(2\tilde{\theta_a})
+\frac{g}{(\pi\alpha)^2}\int \cos (2\tilde{\phi_a})
\end{equation}
Where: $\tilde{\phi_a}=\sqrt{2}\phi_a$ and
$\theta_a=\sqrt{2}\tilde{\theta_a}$.

Introducing the fermion fields
$\Psi_{R,L}=\frac{e^{\imath(\tilde{\theta_a}+\tilde{\phi_a})}}{\sqrt{2\pi
\alpha}}$,
one obtains the fermionized Hamiltonian\cite{shelton_spin_ladders}:

\begin{equation}
H=-\imath v_F \int dx(\Psi^\dagger_R\partial_x\Psi_R-
\Psi^\dagger_L\partial_x\Psi_L) +2 t_\perp \int dx (\Psi^\dagger_R
\Psi^\dagger_L+ \Psi_L\Psi_R) +\frac{g}{\pi\alpha} \int dx
(\Psi^\dagger_R\Psi_L +\Psi^\dagger_L \Psi_R)
\end{equation}

Such Hamiltonian can easily be diagonalized\cite{shelton_spin_ladders}.
It is clear that expectation values such as: $\langle \Psi_R \rangle$
are zero
on the ground state of this Hamiltonian due to the Mermin-Wagner theorem .
This implies $\langle
\sin(\sqrt{2}\phi_a)\cos(\sqrt{2}\theta_a)\rangle=0$.
If one could have $\langle \theta_a^2 \rangle$, $\langle \phi_a^2
\rangle$,$<\infty$ one would expect such quantity to be non-zero.
This indicates that the presence of a supersolid phase in the two
chain system is impossible. In a system of many coupled chains,
however, the Mermin-Wagner does not apply and we cannot rule out a
possible supersolid phase.

\begin{table}\label{4sectors_table}
\caption{the 4 sectors of the two coupled bosonic chains  model}
\begin{tabular}{ccccc}
 &I&II&III&IV\\
\tableline
$K_s$&$<1$&$<1/4$&$>1$&$>1/4$ \\
$K_a$&$<1/2$&$>1/2$&$<1/2$&$>1/2$ \\
\tableline
$\theta_a,\phi_a$&$\langle\phi_a\rangle =\frac \pi {\sqrt{8}}
$&$\langle\theta_a\rangle
=0$&$\langle\phi_a\rangle =\frac \pi {\sqrt{8}}$&$\langle\theta_a\rangle
=0$\\
phase & CDW-  &$CDW ^{4\pi\rho_0}$  & BPSF & SF+\\
Order Parameter&$e^{\imath\sqrt{2}\phi_s}$&$e^{\imath\sqrt{8}\phi_s}$&
$e^{\imath\sqrt{2}\theta_s}$&$e^{\imath\frac{\theta_s}{\sqrt{2}}}$
\end{tabular}
\label{table4}
\end{table}

\begin{figure}
 \centerline{\epsfig{file=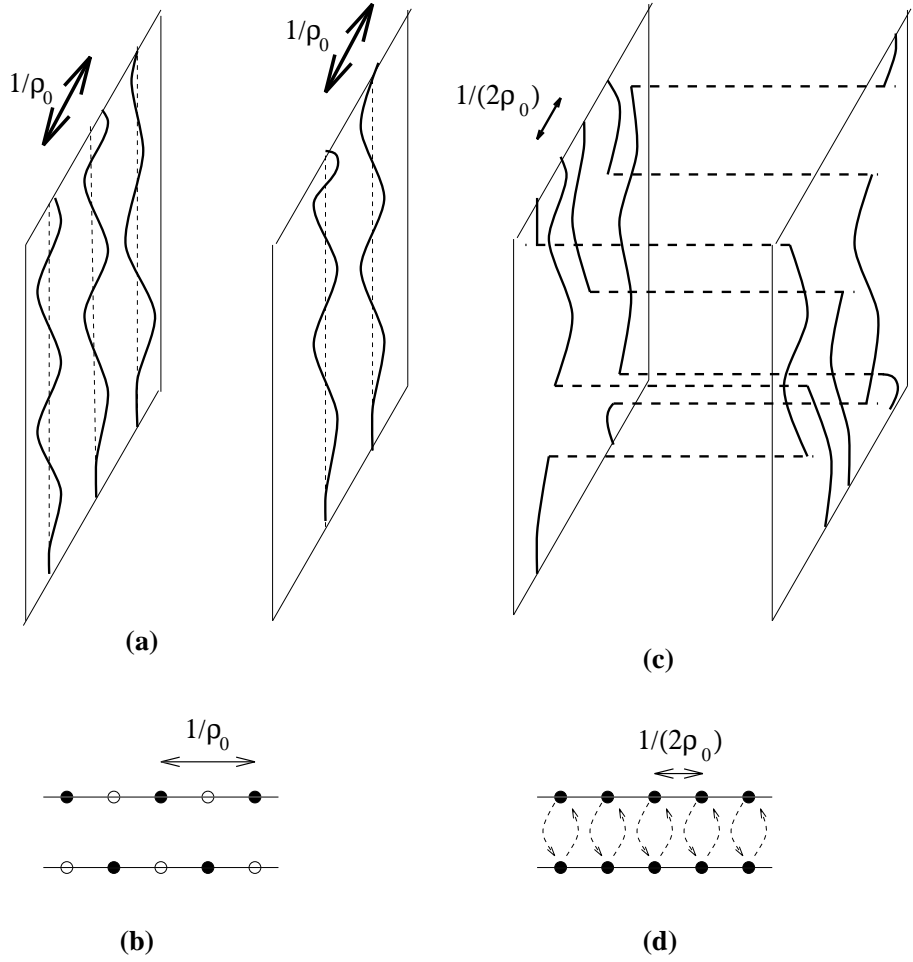,angle=-90,width=12cm}}
   \vspace{0.5cm}
\caption{The behavior of vortices in the $2\pi\rho_0$  solid and
in the $4\pi\rho_0$ solid.
(a) In  the $2\pi\rho_0$ solid, the vortices are confined
in the planes and the average distance between two vortices is
$1/\rho_0$.\\
(b) Top view of the $2\pi\rho_0$ solid.\\
(c) In the $4\pi\rho_0$ solid, vortices are hopping from one plane to
the other, leading to a halving of the distance between two maximums
of the vortex density.\\
(d)  Top view of the $4\pi\rho_0$ solid.\\
}
\label{fig:phases}
\end{figure}

\bibliographystyle{prsty}

\end{document}